\documentclass[a4paper,11pt]{article}
\pdfoutput=1 % if your are submitting a pdflatex (i.e. if you have
\usepackage{jcappub} % for details on the use of the package, please
\usepackage[T1]{fontenc} % if needed
\usepackage{subfigure,float}

\DeclareMathOperator{\arccosh}{arccosh}

\newcommand{\aobs}{\ensuremath{ a_{\mathrm{obs} }}}
\newcommand{\vaobs}{\ensuremath{ \vec{a}_{\mathrm{obs} }}}
\newcommand{\aN}{\ensuremath{ a_{\mathrm{N} }}}
\newcommand{\interp}{\ensuremath{\mathcal{M}}}

\newcommand{\reschifunc}{\ensuremath{ {\cal I}  }}
\newcommand{\ini}{\ensuremath{{\rm in}}}
\newcommand{\chifunc}{\ensuremath{\beta}}
\newcommand{\chifunct}{\ensuremath{\tilde{\chifunc}}}

\newcommand{\abs}[1]{\ensuremath{ \left| #1\right|  } }
\newcommand{\grad}{\ensuremath{\vec{\nabla}}}   
\newcommand{\Jcal}{\ensuremath{{\cal J}}}      
\newcommand{\Scal}{\ensuremath{{\cal S}}}      
\newcommand{\Wcal}{\ensuremath{{\cal W}}}      
\newcommand{\Vcal}{\ensuremath{{\cal V}}}      
      
\newcommand{\Warg}{\ensuremath{u}}      
\newcommand{\zt}{\ensuremath{\tilde{z}}}
      
\newcommand{\Pcal}{\ensuremath{{\cal P}}}      
\newcommand{\Qcal}{\ensuremath{{\cal Q}}}      
\newcommand{\Ycal}{\ensuremath{{\cal Y}}}      
\newcommand{\Fcal}{\ensuremath{{\cal F}}}      
\newcommand{\Kcal}{\ensuremath{{\cal K}}}      
\newcommand{\Lcal}{\ensuremath{{\cal L}}}      
\newcommand{\Ocal}{\ensuremath{{\cal O}}}      
\newcommand{\KB}{\ensuremath{K_{B}}}     
\newcommand{\kB}{\ensuremath{k_{B}}}     
\newcommand{\GN}{\ensuremath{G_{N}}} 
\newcommand{\Phit}{\ensuremath{\tilde{\Phi}}}
\newcommand{\Gt}{\ensuremath{\tilde{G}}}
\newcommand{\Gh}{\ensuremath{G}}
\newcommand{\dt}{\ensuremath{{\rm d}t}}
\newcommand{\dx}{\ensuremath{{\rm d}x}}

\newcommand{\ds}{\ensuremath{{\rm d}s}}
\newcommand{\dr}{\ensuremath{{\rm d}r}}
\newcommand{\dd}{\ensuremath{{\rm d}}}
\newcommand{\rstar}{\ensuremath{R_{{*}}}}
\newcommand{\rI}{\ensuremath{r_{{\rm I}}}}
\newcommand{\rM}{\ensuremath{r_{{\rm M}}}}
\newcommand{\rMh}{\ensuremath{ \hat{r}_{\rm M}  }}
\newcommand{\rhoI}{\ensuremath{\rho_{{\rm I}}}}
\newcommand{\rhoM}{\ensuremath{\rho_{{\rm M}}}}
\newcommand{\rh}{\ensuremath{ \hat{r}}}
\newcommand{\drh}{\ensuremath{{\rm d}\rh}}
\newcommand{\rC}{\ensuremath{r_{{\rm C}}}}

\newcommand{\rhodl}{\ensuremath{\hat{\rho}}}
\newcommand{\Phidl}{\ensuremath{\hat{\Phi}}}
\newcommand{\chidl}{\ensuremath{\hat{\chi}}}
\newcommand{\Phitdl}{\ensuremath{\hat{\Phit}}}
\newcommand{\pPhi}{\ensuremath{P_{\Phi}}}
\newcommand{\pPhidl}{\ensuremath{\hat{P}_{\Phi}}}
\newcommand{\massgas}{\ensuremath{ m_{{\rm gas}} }}
\newcommand{\mugas}{\ensuremath{ \mu_{{\rm gas}} }}
\newcommand{\muAeST}{\ensuremath{\mu}}
\newcommand{\muAeSTs}{\ensuremath{ \tilde{\mu}}}
\newcommand{\muAeSTdl}{\ensuremath{\hat{\mu}}}
\newcommand{\lambdas}{\lambda_s}
\newcommand{\sign}{{\rm Sign}}

\title{Towards galaxy cluster models in Aether-Scalar-Tensor theory: isothermal spheres and curiosities}

\author[a,b]{A. Durakovic,\footnote{Corresponding author.}}
\author[a]{C. Skordis}

\affiliation[a]{CEICO - FZU, Institute of Physics of the Czech Academy of Sciences,\\Na Slovance 1999/2, 182 00 Prague 8, Czechia}
\affiliation[b]{Observatoire astronomique de Strasbourg, Université de Strasbourg,\\11 Rue de l'Université, 67000 Strasbourg,  France}
\emailAdd{amel@fzu.cz}
\emailAdd{skordis@fzu.cz}

\abstract{
The Aether-Scalar-Tensor (AeST) theory is an extension of General Relativity (GR) which can support Modified Newtonian Dynamics (MOND) behaviour in its static
weak-field limit, and cosmological evolution resembling $\Lambda$CDM. 
We consider static spherically symmetric weak-field solutions in this theory and show that the resulting equations can be reduced to a single equation for the gravitational potential.
The reduced equation has apparent isolated singularities when the derivative of the potential passes through zero and we show how these are removed
by evolving, instead, the canonical momentum of the corresponding Hamiltonian system that we find. We construct solutions in three cases: (i) vacuum outside a bounded spherical object, (ii) within an extended prescribed source, and (iii) isothermal gas in hydrostatic equilibrium, serving as a simplified model for galaxy clusters.
We show that the oscillatory regime that follows the Newtonian and MOND regimes, obtained in previous works in the vacuum case, also persists for isothermal spheres, and we show that the gas density profiles in AeST may become more compressed than their Newtonian or MOND counterparts.
We construct the Radial Acceleration Relation (RAR) in AeST for isothermal spheres and find that it can display a peak, an enhancement with respect to the MOND RAR, at an acceleration range determined by the value of the AeST weak-field mass parameter, the mass of the system and the boundary value of the gravitational potential. For lower accelerations, the AeST RAR drops below the MOND expectation, 
as if there is a negative mass density. Similar observational features of the galaxy cluster RAR have been reported. This illustrates the potential of AeST to address the shortcomings of MOND in galaxy clusters, but a full quantitative comparison with observations will require going beyond the isothermal case.
}

\begin{document}
\maketitle
\flushbottom

\section{Introduction}
Galactic and extra-galactic phenomena cannot be accounted for by the dynamics of baryonic matter and radiation (including neutrinos) 
 evolving under gravity described by general relativity (GR). Velocities of stars~\cite{Rubin:1970zza,Rubin:1980zd} and gas~\cite{Bosma:1981zz} in the outskirts of disk galaxies 
are found to asymptote, rather than decline. Assuming GR, or Newtonian gravity for weak fields, relaxed galaxy clusters need more than the baryonic mass to account for the velocity 
dispersion of galaxies~\cite{Zwicky:1933gu} in them, for the thermodynamic state of the gas inferred through X-ray emission~\cite{White:1993wm} and 
the thermal Sunyaev-Zel'dovich effect~\cite{1997ApJ...485....1M,Grego:2000rd}, and for the observed gravitational lensing~\cite{1989ApJ...344..637G}. 
Cosmologically, the observed clustering of galaxies~\cite{2dFGRS:2001csf,Ivanov_2020} and the anisotropies of the Cosmic Microwave Background Radiation are also in stark conflict
with a model of the Universe based on GR containing only baryons and radiation~\cite{Boomerang:2000efg,WMAP:2003ivt,Planck:2018vyg}.

The common approach to resolving all these discrepancies is to posit the existence of a dark matter (DM) component --- a matter component with feeble or no
interactions with baryonic matter other than through gravity. Within the DM paradigm, the most successful (and simplest) model is that of Cold Dark Matter (CDM),
modelled as a collection of particles evolving with the Boltzmann equation. Cosmologically, this leads to a  dust (pressureless matter) 
component which outweighs baryons approximately $5:1$ and forms diverse halos and subhalos around galaxies and galaxy clusters.
DM provides the additional gravitational force necessary in galaxies, and the additional mass to account for cluster observations. 
With the inclusion of a cosmological constant $\Lambda$, the resulting $\Lambda$CDM model is in broad agreement with all cosmological observations at the scale of $\sim Mpc$ or larger.

Direct %\cite{ATLAS:2019wdu,LZ:2022lsv,XENON:2023cxc,ADMX:2021nhd}
 and indirect~\cite{Workman:2022ynf} %\cite{Hess:2021cdp,Hooper:2018kfv}
searches for the dark matter particle have, however, been negative so far,
leaving open the possibility of an extension of GR being responsible for the discrepancies observed.
Moreover, there are regularities in the internal dynamics of galaxies permitting to describe them without a DM component
but by assuming that the observed acceleration $\aobs$ is given by $\aobs = \sqrt{a_0}\sqrt{\aN}$
  when gravitational accelerations are smaller than a threshold $a_0 \sim 1.2\times 10^{-10} m/s^2$ where $\aN$ is
Newtonian acceleration sourced only by the observable baryons.
This is the Modified Newtonian Dynamics (MOND)\cite{Milgrom:1983pn,Milgrom:1983zz,Milgrom:1983ca} proposal.
 Immediate consequences are: the independence of the rotational velocity on the orbital radius,  %$v^2/r = \sqrt{a_{0}} \sqrt{GM_b/r^2} \Rightarrow v^2 = \sqrt{a_0} \sqrt{GM_b} \equiv v^2_{\infty}$, 
and the baryonic Tully-Fisher relation 
%directly follows $v_{\infty}^4 = \left( G a_0 \right) M_b$
% which relates the asymptotic velocity $v_{\infty}$ to \emph{only} the \emph{baryonic} mass 
%of the disk galaxy $M_{b}$, with also has 
which has broad observational support~\cite{Lelli:2019igz}. 

At accelerations larger than $a_0$ Newtonian gravity should be restored.
This is generally achieved through an interpolation function $\interp$ such that $\interp\left(x \right) \aobs = \aN$ where $x= \aobs/a_0$. Thus,
 Newtonian behaviour ensues if $\interp\left(x\right) \to 1$ when $x \gg 1$ while MOND behaviour emerges if $\interp\left(x\right) \to x $ when $x \ll 1$.
%Regularities in galactic 
The modified Poisson equation
\begin{align}
\grad \cdot \left( \interp \; \grad \Phi \right) = 4 \pi \GN \rho_b
\label{eq_MOND}
\end{align}
where $\GN$ is Newton's constant and $\rho_b$ the mass density of baryons,
readily accommodates MOND behaviour as an extension of Newtonian gravity by letting $\vaobs= - \grad \Phi$~\footnote{Additional non-gravitational forces may be present
such that $\vaobs = \vec{a}_{ {\rm grav}} + \vec{a}_{ {\rm non-grav}}$. In that case only $\vec{a}_{ {\rm grav}} \equiv \grad \Phi $ is affected through \eqref{eq_MOND}.}. 
It has a variational formulation, the AQUAL (Aquadratic Lagrangian) non-relativistic gravity  with Lagrangian
$\Lcal = \Jcal(y) + 4\pi G \Phi \rho_b$ where $y\equiv |\nabla \Phi|^2/a_0^2 = x^2$ and $d\Jcal/dy = \interp(x)$, hence, the deep MOND limit 
requires $\Jcal \to y^{3/2} \propto |\nabla \Phi|^3$~\cite{Bekenstein:1984tv}. 
Alternatively, MOND can be seen as a modification of inertia \cite{Milgrom:2022ifm}, though this will not be pursued here.

MOND implies an algebraic relation between the acceleration expected from the baryons only, $\aN$,
 and the observed acceleration $\aobs$. In the context of rotation curves of disk galaxies this relation is known as the Radial Acceleration Relation (RAR) 
and it is observationally supported by the rotation curves of a large sample of disk galaxies \cite{Lelli:2016cui}; see \cite{Desmond:2023wqf}  and \cite{Desmond:2023urj} for recent studies.
The observational support for the RAR has recently been extended by about two orders of magnitude in acceleration using weak lensing \cite{Brouwer:2021nsr} in place of baryonic tracers, 
consistent with the deep MOND scaling of the acceleration $\aobs \sim \sqrt{\aN}$ to a previously unexplored range. It has recently been shown that the RAR is a unique 
fundamental relation in galaxies and that all correlations present in galaxies are manifestations of the RAR~\cite{Stiskalek:2023amy}. 

It is commonly accepted that MOND fails in accounting for the profiles of galaxy clusters \cite{Sanders:2002ue}. The amount of gravitational force needed to support the observed profiles is more than MOND predicts. Radial Acceleration Relations for galaxy clusters have also been constructed using weak and strong lensing \cite{Tian:2020qjd}, galaxy kinematics \cite{Li:2023zua} and a combination of X-ray and thermal Sunyaev-Zel'dovich observations \cite{Eckert:2022chs}. In all cases, the galaxy cluster RAR is above the MOND RAR for disk galaxies. For small enough accelerations, at the outskirts of galaxy clusters, observations suggest that the galaxy cluster RAR even goes below the MOND RAR of disk galaxies. In the context of MOND, this suggests an extra dark component with an apparent negative density in the outskirts.

Extensions of GR, limiting to MOND behaviour in the quasistatic, weak-field regime of galaxies, are necessary if one is to further test this line of investigation 
with gravitational lensing, and with cosmology.
GR extensions along these lines have 
been constructed throughout the years~\cite{Bekenstein:1984tv,Bekenstein:1988zy,Sanders:1996wk,Bekenstein:2004ne,Sanders2005,Skordis:2008pq,Milgrom:2009gv,Blanchet:2012ub}; 
see \cite{Famaey:2011kh} for a review. 
Until recently the most studied candidate has been the Tensor-Vector-Scalar (TeVeS)~\cite{Bekenstein:2004ne}, which, however, is no longer observationally viable on account of 
the CMB anisotropies~\cite{SkordisEtAl2005} and the absence of a significant delay between gravitational waves and the electromagnetic counterpart~\cite{Hou:2018djz,Skordis:2019fxt} which TeVeS predicts.

Using similar field content as in TeVeS, namely a scalar $\phi$ and unit-timelike vector field $A_\mu$, 
the Aether-Scalar-Tensor (AeST) theory has recently been proposed \cite{Skordis:2020eui}. 
It incorporates MOND-like behaviour in its quasi-static limit on small scales, has a $\Lambda$CDM on the largest scales, i.e., effective cosmic dust behaviour 
compatible with observations of the CMB anisotropies and large-scale structure, and has gravitational tensor-mode waves which propagate with the speed of light.
Further studies of the AeST theory have been performed in~\cite{Skordis:2021mry,Bernardo:2022acn,Kashfi:2022dyb,Mistele:2021qvz,Mistele:2023paq,Tian:2023gjt,Llinares:2023lky,Verwayen:2023sds,Mistele:2023fwd,Bataki:2023uuy}.

The phenomenology of the spherically symmetric quasi-static weak-field AeST equations has been studied in \cite{Verwayen:2023sds}  in the case 
of uniform density and Hernquist profile sources as well as in vacuum outside such source profiles. 
In cases which depart from from spherical symmetry, for instance, disk galaxy solutions, the curl of the vector field can also play a role 
and the weak-field AeST equations which include this have also been considered~\cite{Mistele:2023fwd}. Here we extend the previous studies
by investigating the weak-field limit of AeST for spherically symmetric systems, having in mind the application to models of galaxy clusters. 
As such, we consider a simplified model of the galaxy cluster,  the (self-gravitating) hydrostatic isothermal gas sphere. 

The article is organized as follows.
In Section~\ref{sec:theorypres}, we present the AeST action and we derive the weak-field equations in Section~\ref{sec:wkfl}. We then reduce the resulting two-field system to a single equation 
for the gravitational potential in Section~\ref{sec:oneeq}. The reduced equation is a modified Helmholtz equation which includes a mass term in addition to the modified Laplacian term of MOND.
We consider the problem of the apparent singularities due to the oscillatory behaviour that the new mass term induces in Section~\ref{sec:prob}. 
In Section~\ref{sec:hamsys}, we find that the same system can be described in Hamiltonian form in a way which is manifestly singularity free.
 The system can therefore be solved and transformed back to the original variables. 
We illustrate the procedure for the vacuum case in Section~\ref{sec:vac}, the case of a prescribed source in Section~\ref{sec:pres}, 
and finally the case of the hydrostatic isothermal source in Section~\ref{sec:isoth}. In  Section~\ref{sec:rar} we
construct the RAR in  AeST for the case of a hydrostatic isothermal source and uncover novel features which are not present in MOND.

\section{The Aether-Scalar-Tensor theory} 
\subsection{Field content and the theory Lagrangian density}
\label{sec:theorypres}
The field content of AeST is the spacetime metric $g_{\mu \nu}$ (tensor field), a scalar field $\phi$ and a unit time-like vector field $A^\mu$, i.e., one for which $g_{\mu \nu} A^{\mu} A^{\nu} = -1$.
The time-like vector field $A^{\mu}$ allows the construction of kinetic terms not otherwise present in a local, purely metric theory. A projection of the scalar kinetic term onto the preferred time-like direction $A^{\mu}$ gives the scalar
\begin{align}
\Qcal \equiv A^{\mu} \nabla_{\mu} \phi,
\label{def_Qcal}
\end{align}
and a projection onto directions perpendicular to $A^{\mu}$, facilitated by the projector $g^{\mu \nu} + A^{\mu} A^{\nu}$, gives the scalar
\begin{align}
\Ycal \equiv \left(g^{\mu \nu} + A^{\mu} A^{\nu}\right) \nabla_{\mu} \phi \nabla_{\nu} \phi.
\label{def_Ycal}
\end{align}
Another (projected) kinetic term is the acceleration of $A^{\mu}$ which is its gradient projected onto the vector $A^{\mu}$ itself
\begin{align}
J_{\mu} = A^{\alpha} \nabla_{\alpha} A_{\mu}.
\label{def_J}
\end{align}

The total Lagrangian density is given by the sum of the AeST and matter Lagrangian densities $\Lcal_{{\rm AeST}}$ and $\Lcal_{{\rm matter}}$ respectively, that is $\Lcal = \Lcal_{{\rm AeST}} + \Lcal_{{\rm matter}}$.
The AeST Lagrangian density is the Einstein-Hilbert Lagrangian density of $g_{\mu \nu}$, to which matter couples minimally; the Maxwell-like kinetic term $F_{\mu \nu} F^{\mu \nu}$ associated with $A^{\mu}$; 
a coupling between the projected vector gradient $J_{\mu}$ and $\nabla_{\mu} \phi$; and an unspecified function $\Fcal\left(\Ycal, \Qcal \right)$ of the scalars $\Ycal$ and $\Qcal$. 
Explicitly we have
\begin{align}
\Lcal_{{\rm AeST}}= \frac{\sqrt{-g}}{16 \pi \Gt} \left\{ 
R - \frac{\KB}{2} F_{\mu \nu} F^{\mu \nu} +  \left( 2- \KB \right) \left(2J^{\mu} \nabla_{\mu} \phi -  \Fcal\right) - \lambda \left( A^{\mu} A_{\mu} + 1\right)
\right\}
\label{AeST_Lagrangian}
\end{align}
where $\Gt$ is the \emph{bare} gravitational constant, $\KB$ is a coupling constant taking values between $0 < \KB < 2$ \cite{Skordis:2020eui,Skordis:2021mry}, and $\lambda$ is a Lagrange multiplier field that enforces the unit time-like constraint $g_{\mu \nu} A^{\mu} A^{\nu} = -1$.

\subsection{Forms of the free function $\Fcal$}
The form of $\Fcal$ determines the dynamics of AeST in a cosmological setting and the effective profiles of the gravitational potential in the quasistatic weak-field limit.
In order to recover $\Lambda$CDM behaviour, $\Fcal$ has features akin to shift-symmetric K-essence~\cite{Scherrer:2004au} which is the low energy limit of
 ghost condensation~\cite{ArkaniHamedEtAl2003,Arkani-Hamed:2005teg}. This means that on a homogeneous-isotropic spacetime, where $\Ycal=0$, we may define 
the function $\Kcal(\Qcal) \equiv  -\frac{1}{2} \Fcal(0,\Qcal)$ which is required to have a Taylor expansion of the form $\Kcal(\Qcal) =  \Kcal_2 \left( \Qcal - \Qcal_0   \right)^2 + \ldots$
around a non-zero constant $\Qcal_0$, while $\Kcal_2$ is another constant
and where $(\ldots)$ denote higher order terms in this expansion.
Assuming a synchronous coordinate system where $g_{00} = -1$,
this ensures that cosmologically $\Qcal$ evolves with redshift $z$ as $\Qcal = \Qcal_0 + I_0(1+z)^3 + \ldots$ where $(\ldots)$ denote higher powers of $1+z$ which are relevant in the early Universe,
and $I_0$ is an initial condition which sets the initial displacement of $\Qcal$ away from its minimum at $\Qcal_0$.
These considerations on $\Kcal$ (and thus on $\Fcal$), result in AeST providing cosmological energy density scaling as $(1+z)^3$ plus corrections scaling with higher powers of $1+z$, that is,
approximately that of dust.
The evolution of $\Qcal$ initially displaced away from $\Qcal_0$ is to tend towards it, and the cosmological dust density $\Omega_{{\rm AeST}}$
is set by the initial displacement of $\Qcal$ from $\Qcal_0$~\cite{Skordis:2020eui}.

On quasistatic backgrounds in the late Universe, $\Qcal$ may be assumed to be sufficiently close to its minimum up to small fluctuations. This introduces another reduction of $\Fcal$,
the function $\Jcal(\Ycal) \equiv \Fcal(\Ycal,\Qcal_0)/(2-\KB)$, which is subsequently found to appear in the modified Poisson equation and must be appropriately chosen to lead to MOND
and to Newton under appropriate conditions. 
The former (MOND) can happen provided 
\begin{align}
\Jcal(\Ycal) =  \frac{2\lambdas}{3(1+\lambdas) a_0} |\Ycal|^{3/2}   + \Ocal(\Ycal^2)
\label{Jcal_MOND_limit}
\end{align}
as $\Ycal\rightarrow 0$, (that is, when $\sqrt{\Ycal} \ll a_0$) which captures the AQUAL requirement $ |\Ycal|^{3/2}$ (see below).
Newtonian behaviour results if
\begin{align}
\Jcal(\Ycal) = \lambdas \Ycal + \ldots
\label{def_lambdas}
\end{align}
as $\Ycal\rightarrow \infty$ (that is, when $\sqrt{\Ycal} \gg a_0$), where $(\ldots)$  denote subleading terms to $\Ycal$. Here $\lambdas > 0$ is a free parameter  defined to be part of $\Fcal$; see~\cite{Skordis:2021mry}. The effect of $\lambdas$ is to completely screen the MOND contributions in the large gradient limit as $\lambdas\rightarrow \infty$.

With the above considerations we assume that $\Fcal$ consists of two pieces $\Kcal$ and $\Jcal$, and that the former has the Taylor expansion discussed above. Then,
\begin{align}
\Fcal\left(\Ycal,\Qcal\right) = - 2 \Kcal_2 \Qcal_0^2  \left(\Qcal - \Qcal_0\right)^2 + \Jcal(\Ycal).
\end{align}

The exact form of $\Jcal(\Ycal )$ determines the MOND interpolation function $\interp(x)$ and as we show below, in spherically symmetric situations, or when curls are subdominant,
the simple choice of 
\begin{align}
\Jcal(\Ycal) = \frac{2 \Ycal^{3/2}}{3 a_0} 
\label{eq:simpcho}
\end{align}
that we referred to as ``totally screened'' interpolation function, corresponds to the MOND interpolation function
\begin{align}
\interp(x) = \frac{\sqrt{1+4x}-1}{\sqrt{1+4x}+1}
\label{eq:simple_interp}
\end{align}
where $x \equiv \left|\nabla \Phi\right|/a_0$, in the modified Poisson equation
\begin{align}
\nabla \cdot \left[ \interp(x) \nabla \Phi \right] + \ldots = 4 \pi G \rho_b.
\end{align}
The function $\interp$ in \eqref{eq:simple_interp} has the expected limits $\interp(x) \to x$ for $x\ll 1$ (deep MOND regime) and the Newtonian limit $\interp(x) \to 1$ when $x \gg 1$.
The exact relation between the function $\Jcal$ in the Lagrangian density and the MOND interpolation function $\interp$ is made explicit in Section~\ref{sec:intpf} and Section~\ref{sec:oneeq}.
We refer to Appendix~\ref{sec:screening} for an example of a choice of $\Jcal\left(\Ycal\right)$ for the case of the MOND interpolation function $\interp(x)=x/\left(x+1\right)$. 
Here, we keep to the  totally screened  choice of \eqref{eq:simpcho} when discussing our numerical results for the remainder of the article. A generalization of this 
is found in Appendix~\ref{sec:screening}.

\subsection{The weak-field, quasi-static limit of AeST}
\label{sec:wkfl}

\subsubsection{Weak-field limit and the quasistatic Lagrangian density}
We are interested in the weak-field quasistatic limit of AeST applied to the case of spherical symmetry in order to eventually build simple models of galaxy clusters.
We assume that the spacetime is approximately flat, plus small metric fluctuations parametrised in the Newtonian gauge by the 
 potentials $\Psi$ and $\Phi$ such that
\begin{align}
\ds^2 = - \left(1+2 \Psi \right) \dt^2 + \left(1-2 \Phi\right) \gamma_{ij} \dx^{i} \dx^j
\end{align}
where $\gamma_{ij}$ is the flat Euclidean metric in arbitrary spatial coordinates.

The unit time-like constraint $g_{\mu \nu} A^{\mu} A^{\nu} = -1$ implies that
\begin{align}
A_{0} &= -\sqrt{1+2 \Psi} \sqrt{{1+ \left(1 - 2 \Phi\right)^{-1} |\mathbf{A}|^2}} = -1 - \Psi + \frac{\Psi^2}{2} - \frac{|\vec{A}|^2}{2} + \ldots.
\end{align}
and the spatial part of $A_\mu$ denoted $\vec{A}$, is decomposed into the gradient of a scalar $\alpha$ and a curl $\vec{\alpha}_{\perp}$ so that 
\begin{align}
\vec{A} = \grad \alpha + \vec{h}
\end{align}
 with  $\grad \cdot \vec{h} = 0$. In the following we set $\vec{h} = 0$ as it is not relevant to the spherically symmetric situations we study, 
and is expected to be subdominant even in cases corresponding to disk galaxies; see for example \cite{Mistele:2023fwd}. 
We assume that the quasistatic systems we consider are formed in the late universe where the cosmological contribution of AeST has settled very close to the minimum of the function $\Kcal$, that is $\Qcal \rightarrow \Qcal_0$, up to small fluctuations. Then we may expand $\phi$ about this minimum so that
\begin{align}
\phi = \Qcal_0 t + \varphi.
\end{align}
Finally, we impose that all variables are time independent so that $\dot{\Psi}=\dot{\Phi}=\dot{\alpha}=\dot{\varphi}=0$.

Defining the gauge invariant variable $\chi \equiv \varphi + \Qcal_{0} \alpha$  (see \cite{Skordis:2021mry}) that combines the scalar perturbation $\varphi$ and 
the gradient mode $\alpha$, the (spatial) kinetic scalar $\Ycal$ from \eqref{def_Ycal} equals
\begin{align}
\Ycal &= \left( g^{\mu \nu} + A^{\mu}A^{\nu} \right) \nabla_{\mu} \left(\Qcal_{0}t + \varphi \right) \nabla_\nu \left(\Qcal_{0}t + \varphi \right) = |\grad \chi|^2.
\end{align}
Hence $\Jcal\left( \Ycal\right) = 2 \Ycal^{3/2}/3 a_0$ in \eqref{eq:simpcho} becomes $\Jcal\left(\Ycal \right) = 2|\grad \chi|^3/3 a_0$.

Let us now consider the Lagrangian density \eqref{AeST_Lagrangian} and expand the various terms.
The Ricci scalar itself only involves the metric potentials $\Psi$ and $\Phi$, and after partial integrations which remove second spatial derivatives we find
$\sqrt{-g} R = 2|\grad \Phi|^2 - 4\grad \Phi \cdot \grad \Psi$.
From the vector field kinetic term comes only $F_{\mu \nu} F^{\mu \nu} = - 2 |\grad \Psi|^2$
and from the vector-scalar coupling $J^{\mu} \nabla_\mu \phi =  \grad \Psi \cdot \grad \chi$.
The free function $\Fcal$ evaluates to $\Fcal=  - 2 \Kcal_2  \Qcal_0^2  \Psi^2  + \Jcal $.
To calculate the contribution from the matter Lagrangian 
we set $T_{00} = \rho_b$ to be the baryon density, while $T_{0i} = T_{ij} = 0$, 
and use $-2 \delta (\sqrt{-g} \Lcal_{{\rm matter}} ) = \sqrt{-g} T_{\mu\nu} \delta g^{\mu\nu} =  2\rho_b \Psi$.

Inserting all these terms into \eqref{AeST_Lagrangian} leads to
\begin{align}
\Lcal^{{\rm Quasistatic}}_{{\rm AeST}}  = &2 | \grad \Phi|^2 - 4 \grad \Phi \cdot \grad \Psi + \KB |\grad \Psi|^2 + 2\left( 2- \KB \right) \grad \Psi \cdot \grad \chi
 \nonumber 
\\ &
- \left(2 - \KB \right) \left(  |\grad \chi|^2 + \Jcal \right) +  2 \Kcal_2 \Qcal_0^2 \Psi^2 
- 16 \pi \Gt \rho_b \Psi
\end{align}
where we have multiplied through by $16 \pi \Gt$.  The variation of $\Phi$ gives the condition $\Phi = \Psi$, which is substituted back to give
\begin{align}
-\frac{1}{2-\KB} \Lcal^{{\rm Quasistatic}}_{{\rm AeST}} = & 
  | \grad \Phi|^2 - 2 \grad \Phi \cdot \grad \chi  + |\grad \chi|^2 + \Jcal 
- \muAeST^2 \Phi^2 
+ 8 \pi \Gh \rho_b \Phi 
\label{eq:tlag}
\end{align}
where we have defined the mass scale
\begin{align}
\muAeST^2 \equiv \frac{2 \Kcal_2 \Qcal_0^2}{2-\KB}
\end{align}
and the rescaled gravitational constant $\Gh$ as
\begin{align}
  \Gh \equiv \frac{2\Gt}{2 - \KB}.
\end{align}

\subsubsection{The weak-field system of equations}
\label{sec:intpf}
The resulting equations, obtained by variations of \eqref{eq:tlag} with respect to $\Phi$ and $\chi$, respectively, are
\begin{align}
 \grad^2 \Phit + \muAeST^2 \Phi =& 4 \pi \Gh  \rho_b 
\label{eq_Phit}
\\
\grad^2 \Phit  =& \grad \cdot \left( \Jcal_{\Ycal}  \grad \chi \right) 
\label{eq_chi}
\end{align}
where we have defined the potential $\Phit$ through
\begin{align}
 \Phit \equiv \Phi - \chi
\end{align}
and where $\Jcal_\Ycal \equiv d\Jcal/d\Ycal$. 
The significance of the $\Phit$ combination is that as long as the term proportional to $\muAeST$ is small, $\Phit$ is the usual potential obtained by solving the Poisson equation sourced only by baryons. Starting from \eqref{eq_Phit} and \eqref{eq_chi}, we obtain solutions for $\Phit$ and $\chi$ by casting $\Phi = \Phit + \chi$ and this last equality is used to determine
$\Phi$ from which particle accelerations $\vec{a}$ (and trajectories) are calculated via $ \vec{a} = - \grad \Phi$.
We refer to the set \eqref{eq_Phit} and \eqref{eq_chi} as the two-component system of equations. 

Let us now return to the MOND and Newtonian limits. For this it is sufficient to set $\mu\rightarrow0$. If $\Jcal = \lambdas \Ycal$ (Newtonian limit) according to \eqref{def_lambdas},
 then \eqref{eq_chi} results in $\grad^2 \Phit  =  \lambdas \grad^2 \chi$ and so up to curl modes which are subdominant and unimportant in spherical symmetry, $\chi = \Phit / \lambdas$, so that
\eqref{eq_Phit} turns into a Poisson equation for $\Phi$, i.e. $\grad^2 \Phi = 4 \pi \GN \rho$ with a rescaled Newton's constant corresponding to its measured value as
\begin{align}
\GN &\equiv \frac{1+\lambdas}{\lambdas}  \Gh = \frac{1+\lambdas}{\lambdas} \frac{2\Gt}{2-\KB}.
\label{def_GN}
\end{align}
With the above definition of $\GN$, it can be easily checked that  $ \Jcal = \frac{2\lambdas}{3(1+\lambdas) a_0} |\Ycal|^{3/2}$ according to \eqref{Jcal_MOND_limit}
corresponds to a deep MOND limit.

Notice that as $\lambdas\rightarrow \infty$ in our short discussion above, we have that $\chi \rightarrow 0$ so that $\Phit \rightarrow \Phi$ and $\GN \rightarrow G$. In this sense,
the effects of $\chi$ are totally screened and so we refer to $\lambdas$ as the screening parameter. In practice, $\Jcal$ is not exactly equal to $\lambdas \Ycal$, but contains
subdominant terms in $\Ycal$ (such as $\sqrt{\Ycal}$, $1/\Ycal$ or $\log(\Ycal)$). Therefore, screening is not perfect, in the sense that $\chi'$ is not exactly zero but
 contains terms scaling as $r^n$ with $n>-2$, so that $\Phi'$ always dominates as $r\rightarrow 0$; it is in the limit $r\rightarrow 0$ that screening becomes exact.
Thus the $\lambdas\rightarrow \infty$ limit is to be taken that $\Jcal$ does not contain a term proportional to $\Ycal$ in this limit. We refer the reader 
to Appendix~\ref{sec:screening} for a more detailed discussion.

\subsubsection{Reduction to one equation}
 \label{sec:oneeq}
It is possible to reduce the two-component system of equations \eqref{eq_Phit} and \eqref{eq_chi} 
 to one PDE of the gravitational potential $\Phi$ provided that curl components resulting after integrating  \eqref{eq_chi} are small.

Defining  
\begin{align}
\chifunc \equiv 1+ \Jcal_\Ycal
\label{chifunc}
\end{align}
where  $\chifunc = \chifunc\left(  |\nabla \chi|/a_0 \right)$, the  system  \eqref{eq_Phit} and \eqref{eq_chi} becomes
\begin{align}
\grad^2 \Phi - \grad^2 \chi + \muAeST^2 \Phi  &= 4 \pi \Gh \rho_b 
\label{eq:phieq}
 \\
\grad \cdot \left( \chifunc \grad \chi \right)  &= \grad^2 \Phi.
\label{eq:chieq}
\end{align}
Integrating the last equation, \eqref{eq:chieq}, leads to
\begin{align}
 \chifunc \grad \chi  &= \grad \Phi + \grad \times \vec{h}
 \label{eq:betcase}
\end{align}
where $\grad \times \vec{h}$ is a curl mode. Thus provided that boundary conditions allow $\vec{h}$ to be small, we can eliminate $\chi$ in favour of $\Phi$. 
Then we have that
\begin{align}
\chifunc  |\grad \chi| = |\grad \Phi|
\end{align}
and, after defining $y \equiv |\grad \chi|/a_0$ and $x \equiv |\grad \Phi|/a_0$, we define a new function $\chifunct$  through
\begin{align}
\chifunct\left(y \right) \equiv   \chifunc \left( y\right) y  = x.
\label{eq:betp}
\end{align}
We are after $y$, so isolating it gives
\begin{align}
y = \chifunct^{-1}(x) 
\label{eq:xby}
\end{align}
and substituting back into \eqref{eq:betcase} (with $\vec{h} = 0$) gives
\begin{align}
\grad \chi = \chifunct^{-1}(x) \frac{\grad \Phi}{ |\grad \Phi|} \equiv \reschifunc(x) \grad \Phi
\label{eq:expchi} 
\end{align}
where $\reschifunc(x) \equiv \chifunct^{-1}(x)/x$. 
Taking the divergence of \eqref{eq:expchi} results in
\begin{align}
\grad^2 \chi = \grad \cdot \left( \reschifunc  \grad \Phi \right)
\end{align}
which allows us to eliminate $\grad^2 \chi$ from \eqref{eq:phieq}  to get the final single-field equation 
\begin{align}
 \grad \cdot \left( \interp\grad \Phi \right) +   \muAeSTs^2 \Phi  = 4 \pi \GN \rho_b.
\label{eq:mastereq}
\end{align}
In getting \eqref{eq:mastereq}  we have defined the $\interp  =\interp(x)$  via
\begin{align}
 \interp  \equiv  \left(1+\beta_0\right) \left(  1 - \reschifunc \right),
\label{eq:em}
\end{align}
the rescaled mass parameter as
\begin{align}
  \muAeSTs^2 \equiv& \left(1 + \beta_0\right) \muAeST^2
\label{eq:mu_resc}
\end{align}
and the inverse screening parameter as
\begin{align}
\beta_0 \equiv& \frac{1}{\lambdas}.
\label{eq:beta_0}
\end{align}
We recognise \eqref{eq:mastereq} to be the AQUAL equation \eqref{eq_MOND} of \cite{Bekenstein:1984tv}, with the additional term $\muAeST^2 \Phi$, turning it into 
a modified Helmholtz equation. The new term $\muAeST^2 \Phi$ becomes important at very large distances away from sources and introduces novel oscillatory features in the solutions
which manifest after the MOND limit occurs~\cite{Verwayen:2023sds}.

\subsubsection{The interpolating function}
Consider the simple function \eqref{eq:simpcho}, which implies that $\lambdas= \infty$ (and hence $\beta_0=0$).  We first compute  
\begin{align}
\chifunc(y) = 1+ \Jcal' =  1 + y,
\end{align}
from which we find $x(y)$ by forming   $\chifunct(y) = y \chifunc =  y+y^2 = x$  from \eqref{eq:betp}, leading to
 \begin{align}
y = \chifunct^{-1}(x) = \frac{ -1 + \sqrt{1+4x} }{2}.
 \end{align}
This in turn gives $\reschifunc(x)  =\chifunct^{-1}(x)/x = ( -1 + \sqrt{1+4x} ) /2x.$ from which we get the 
 MOND interpolating function using \eqref{eq:em}
\begin{align}
\interp(x) = 1 - \reschifunc(x) = \frac{-1 + \sqrt{1+4x}}{ 1 + \sqrt{1+4 x}}. 
\label{eq:intpfunc}
\end{align}
One can easily check that it has the correct limits: when $x \gg 1$ we have that $\interp(x) \to 1$ which is the large gradient limit  giving Newtonian behaviour, 
and when $x \ll 1$ we have that  $\interp\left(x\right) \to x$ leading to the low gradient limit which contains  MOND when $\mu^2 \rightarrow 0$.

In Appendix~\ref{sec:screening}  we compute a more general interpolation function and show that it coincides with \eqref{eq:intpfunc} as $\lambdas\rightarrow \infty$ implying full
consistency. 

\subsubsection{Solving the modified Helmholtz equation in spherical symmetry}
\label{sec:prob}
We are interested in solving \eqref{eq:mastereq} in the case of spherical symmetry, for a variety of source profiles $\rho_b(r)$, including the isothermal case. In spherical symmetry,
\eqref{eq:mastereq}  becomes
\begin{align}
\frac{1}{r^2} \frac{\dd}{\dr} \left( r^2 \interp  \frac{\dd\Phi}{\dr} \right) + \muAeSTs^2 \Phi  = 4 \pi \GN \rho_b
\label{eq:mastereq_spherical}
\end{align}
where $\rho_b = \rho_b(r)$ and $x = |\Phi'|/ a_0$, $x$ being the argument of $\interp$ and primes denoting derivatives w.r.t. $r$. 

In the large gradient limit where  $\interp \rightarrow 1$, \eqref{eq:mastereq_spherical} turns into the Helmholtz equation which 
has exact oscillatory solutions for any source $\rho_b(r)$ given by the homogeneous solution
 $\Phi(r) = C_1 \cos \left( \muAeSTs r \right)/r + C_2 \sin \left( \muAeSTs r \right)/  2 \muAeSTs r$. For more general $\interp$, the oscillations persist but an analytic 
solution is no longer possible. Numerical solutions to the two-component system, which as we have  shown above are equivalent to \eqref{eq:mastereq_spherical}, 
have been studied thoroughly in~\cite{Verwayen:2023sds}; see also \cite{Mistele:2023fwd}. Assuming a source of mass $M$ and that an interpolation function $\interp$ is chosen with the 
correct small gradient limit in order to lead to MOND when $\mu$ is negligible (which happens at a scale 
\begin{align}
\rM \equiv \sqrt{\frac{\GN M}{a_0}},
\label{eq_rM}
\end{align} 
then oscillations begin~\cite{Verwayen:2023sds} at a scale $\rC$ given by
\begin{align}
%\rC = \left(\frac{\rM}{\mu^2}\right)^{1/3}
\rC \approx& \frac{1}{3} \left(\frac{18 \rMh}{\mu^2} \frac{1}{ 1  + 3 |\Delta|  } \right)^{1/3}
\label{eq_rC}
\end{align}
where $\rMh = \lambdas \rM /(1+\lambdas)$ and $\Delta$ is a parameter related to the boundary condition which sets $\Phi$ on the source surface $\rstar$.
The above equation holds if the size of the source $\rstar$ is smaller than $\rM$ and so, the oscillations of $\Phi$ (and also $\chi$) are occurring in vacuum.
 We show below that oscillations persist even if the source is extended, such as the case of an
 isothermal source, however, the scale $\rC$ is modified and we have been unable to find an analytic estimate as in \eqref{eq_rC} in that case.

In practice, oscillations in \eqref{eq:mastereq} for a general $\interp$ cause problems because the derivative from the divergence acts on $|\Phi'|$.
 Since $\Phi$ oscillates it is unavoidable that $|\Phi'| = 0$, introducing a singularity when solving for $\Phi''$ in \eqref{eq:mastereq_spherical}, which we shall see is only apparent.

In the next section we demonstrate that one can find a reduced Lagrangian from which  \eqref{eq:mastereq_spherical} is derived by a variational principle. Then,
one can find the Hamiltonian and recast \eqref{eq:mastereq_spherical} into a pair of (first order) Hamilton's equations for $\Phi$ and its canonical momentum $\pPhi$.
The Hamiltonian system  does not exhibit these apparent singularities and integration can proceed smoothly. 

When an equivalent Hamiltonian system is not known, then it is better to use the two-component system of equations rather than \eqref{eq:mastereq_spherical}, which in 
spherical symmetry are 
\begin{align}
\frac{1}{r^2} \frac{\dd}{\dr} \left[ r^2 (\Phi' - \chi')\right] + \muAeSTs^2 \Phi  &= 4 \pi \GN \rho_b
 \label{eq:chieq2} 
\\
\chifunc \frac{\dd\chi}{\dr}  &=  \Phi'
\label{eq:candiv}.
\end{align}
Details of the setup for the numerical solution of \eqref{eq:chieq2}-\eqref{eq:candiv} are found in Appendix~\ref{sec:twofc}.

\section{Reduced Hamiltonian systems}
\label{sec:hamsys}
%\begin{align}
%L(\Phi, \Phi') = r^2  \left( a_0^2 \Qcal(|\Phi'|/a_0) - \frac{\muAeST^2}{2}\Phi^2 \right)
%\end{align}
%\begin{align}
%L\left(\Phi, \Phi' \right) = r^2 a_0^2 \Qcal(|\Phi'|/a_0) - \frac{ \muAeST^2 r^2 }{2}\Phi^2 + 4\pi G r^2 \rho_b \Phi
%\end{align}
%\begin{align}
%L =  a^2_{0} r^2 \Qcal(|\Phi'|/a_0) - \frac{ \muAeST^2 r^2}{2} \Phi^2 - 4 \pi G \rho_0 \xi^{-1} r^2 \exp\left(-\xi \left( \Phi - \Phi_0 \right) \right)
%\end{align}

\subsection{The master Lagrangian}
Consider the Lagrangian 
\begin{align}
L\left(\Phi, \Phi', r \right) = r^2 a_0^2 \Pcal - \frac{\muAeSTs^2 r^2 }{2}\Phi^2 + 4\pi \GN r^2 \Scal(\Phi,r)
\label{master_L}
\end{align}
where $\Scal = \Scal(\Phi,r)$ is an unspecified function and $\Pcal = \Pcal(x)$ is defined by
\begin{align}
\Pcal(x) \equiv \int^x \dx' \, \interp\left(x'\right) x',
\end{align}
with $x \equiv |\Phi'| / a_0$ as before and $\interp(x)$ a given interpolation function. The Euler-Lagrange equations resulting from  \eqref{master_L} lead directly to
\eqref{eq:mastereq_spherical}, with the density given by
\begin{align}
\rho_b =  \frac{\partial\Scal}{\partial\Phi}.
\end{align}
In order to resolve the issues with the apparent singularities appearing in \eqref{eq:mastereq_spherical} we choose to work in phase space coordinatised by the potential $\Phi$ 
and its canonical momentum $\pPhi$. We find that
\begin{align}
\pPhi \equiv \frac{\partial L}{\partial \Phi'} =& r^2 a_0^2 \frac{d\Pcal }{d\Phi'} = r^2 \interp  \Phi'
\label{P_Phi}
\end{align}
which, defining $z \equiv |\pPhi|/a_0$, may be inverted to give
\begin{align}
\zt \equiv&  \frac{z}{r^2} = x \interp  \qquad \Rightarrow x = x(\zt)
\label{eq_x_zt}
\end{align}
so that
\begin{align}
\Phi' = \frac{x}{z} \pPhi. \label{eq:phiasp}
\end{align}
Then the corresponding Hamiltonian $H \equiv \Phi' \pPhi   - L$ is
\begin{align}
H(\Phi,\pPhi,r) =  \frac{x}{z} \pPhi^2 -  a_0^2 z x + r^2 a_0^2 \int^{\zt} d\zt' x
 + \frac{\muAeSTs^2 r^2 }{2}\Phi^2 - 4\pi \GN r^2 \Scal.
\end{align}
From Hamilton's equations we then find the equations of motion as
\begin{align}
\Phi' =&   \sign(\pPhi) a_0 x
\label{eq_Phi_prime}
\\
 \pPhi' =&  r^2 \left( - \muAeSTs^2 \Phi + 4\pi \GN  \frac{\partial \Scal }{\partial \Phi} \right)
\label{eq_P_Phi_prime}
\end{align}
where $x = x(\zt)$ in \eqref{eq_Phi_prime} is evaluated from \eqref{eq_x_zt}. Note that \eqref{eq_Phi_prime} is consistent with \eqref{P_Phi} and becomes trivial when 
substituting $x = |\Phi'|/a_0$. One can now solve the first-order system \eqref{eq_Phi_prime} and \eqref{eq_P_Phi_prime} or form the second-order ODE for $\pPhi$ as
\begin{align}
 \pPhi'' =&  \frac{2}{r} \pPhi'
+ r^2 \left[  \left( 4\pi \GN  \frac{\partial^2 \Scal }{\partial \Phi^2} - \muAeSTs^2\right) \sign(\pPhi) a_0  x 
 + 4\pi \GN  \frac{\partial^2 \Scal }{\partial \Phi \partial r} 
 \right].
\label{eq_P_Phi_prime_prime}
\end{align}
Notice that no apparent singularities occur when the Hamiltonian formulation is used. Let us also note that for our simple choice of interpolation function \eqref{eq:intpfunc} 
 we have that
\begin{align}
x = \sqrt{\zt} + \zt.
\label{x_zt}
\end{align}
Hence in that case we find
\begin{align}
 \pPhi'' =&  \frac{2}{r} \pPhi'
+ r \left[  \left(  4\pi \GN  \frac{\partial^2 \Scal }{\partial \Phi^2} - \muAeSTs^2\right) \left( \sign(\pPhi) \sqrt{ a_0 |\pPhi|  } + \frac{1}{r} \pPhi  \right)
 + 4\pi \GN r  \frac{\partial^2 \Scal }{\partial \Phi \partial r}
 \right].
\label{eq_P_Phi_prime_prime_func}
\end{align}

\subsection{Vacuum case: $\Scal=0$}
\subsubsection{Vacuum equations and dimensionless variables}
\label{sec:vac}
Outside a source, we may impose the vacuum condition for which $\Scal = 0$. Then the system  \eqref{eq_Phi_prime} and \eqref{eq_P_Phi_prime}  becomes
\begin{align}
\Phi' =&  \frac{ \sign(\pPhi)}{r} \sqrt{a_0 |\pPhi|   } + \frac{1}{r^2} \pPhi 
\label{eq_Phi_prime_vac}
\\
 \pPhi' =&  -r^2  \muAeSTs^2 \Phi 
\label{eq_P_Phi_prime_vac}
\end{align}
and equivalently, the second order equation \eqref{eq_P_Phi_prime_prime_func} becomes
\begin{align}
 \pPhi'' =&  \frac{2}{r} \pPhi' -   \muAeSTs^2 \left[r \, \sign(\pPhi) \sqrt{ a_0 |\pPhi|  } +  \pPhi  \right].
\label{eq_P_Phi_prime_prime_vac}
\end{align}
To solve the above equation we impose boundary conditions translated from the original problem to this one using the definition of $\pPhi$ \eqref{P_Phi} 
and its evolution equation \eqref{eq_P_Phi_prime_vac} as
\begin{align}
\pPhi(r_0) &= r_0^2 \interp\left(|\Phi'(r_0)|/a_0\right) \Phi'(r_0) 
\label{eq:inic1} 
\\
\pPhi'(r_0) &= - r_0^2 \muAeSTs^2 \Phi\left(r_0 \right) 
\label{eq:inic2}.
\end{align}
In the $\mu=0$ case, we can readily solve the above system to get that $\pPhi$ is a constant from \eqref{eq_P_Phi_prime_vac}, so that the force $\Phi'$ becomes a sum of a Newtonian force
$1/r^2$ and a MOND force $1/r$. If $\mu\ne 0$ we cannot find analytic solutions and a numerical implementation must be employed. We leave the details
for  Appendix~\ref{sec:redone} and here we present our results.

%\begin{align}
%\frac{-1 + \sqrt{1+4x}}{ 1 + \sqrt{1+4 x}} x  = \frac{C a_0}{r^2}
%\end{align}

It is easier to work with dimensionless variables adapted to our problem. 
For the case of the point source with mass $M$, a good choice is the MOND radius $\rM$ defined by \eqref{eq_rM} so that the dimensionless radius is $\rh = r/\rM$. 
Then $\rh=1$ is the turning radius from Newtonian behaviour to MOND so that radii larger that $\rh=1$ are in the low gradient regime.
Likewise we define the dimensionless mass parameter $\muAeSTdl^2 \equiv \muAeSTs^2 \rM^2$ and the dimensionless canonical momentum $\pPhidl =   \pPhi/ \GN M $ and 
rescaled potential $\Phidl = \Phi/\sqrt{\GN M a_0}$. In these units, \eqref{eq_Phi_prime_vac} and \eqref{eq_P_Phi_prime_vac} become
\begin{align}
 \frac{\dd\Phidl}{\drh}  =& \frac{\sign(\pPhidl)}{\rh} \sqrt{|\pPhidl|   } + \frac{1}{\rh^2} \pPhidl
\label{eq_Phi_prime_vac_dl}
\\
 \frac{\dd\pPhidl}{\drh} =&  - \rh^2  \muAeSTdl^2 \Phi
\label{eq_P_Phi_prime_vac_dl}
\end{align}
and equivalently
\begin{align}
 \frac{\dd^2\pPhidl}{\drh^2}  =&  \frac{2}{\rh} \frac{\dd\pPhidl}{\drh}   -   \muAeSTdl^2 \left[ \rh \, \sign(\pPhi) \sqrt{ |\pPhidl|  } +  \pPhidl  \right].
\label{eq_P_Phi_prime_prime_vac_dl}
\end{align}
\emph{Note} that this means that the case of a galaxy and a galaxy cluster are cases with \emph{different} $\muAeSTdl$ as their MOND radii are \emph{different}. 
For the point source the parameter $\muAeSTdl^2$ 
is proportional to the mass of the system. 

\subsubsection{Solutions}
\begin{figure}[H]
\centering
\includegraphics[width=0.49\textwidth]{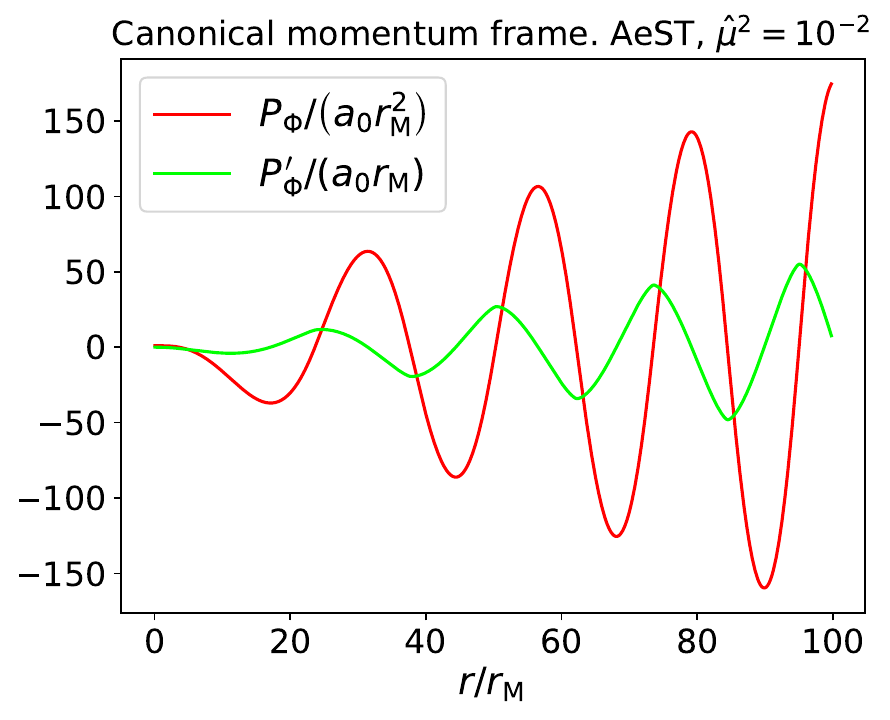}
\includegraphics[width=0.49\textwidth]{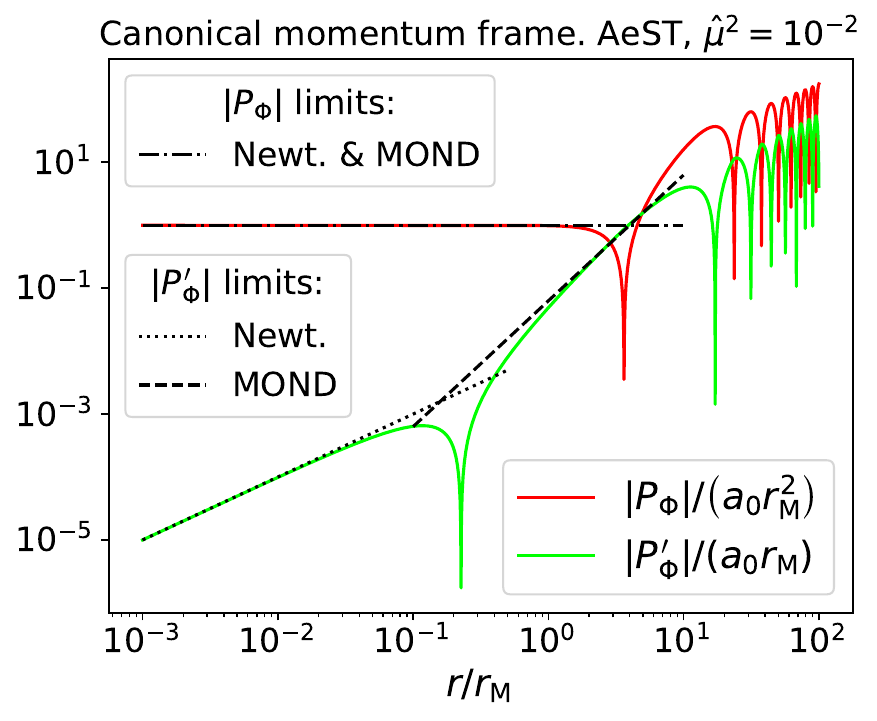}
\\
\includegraphics[width=0.49\textwidth]{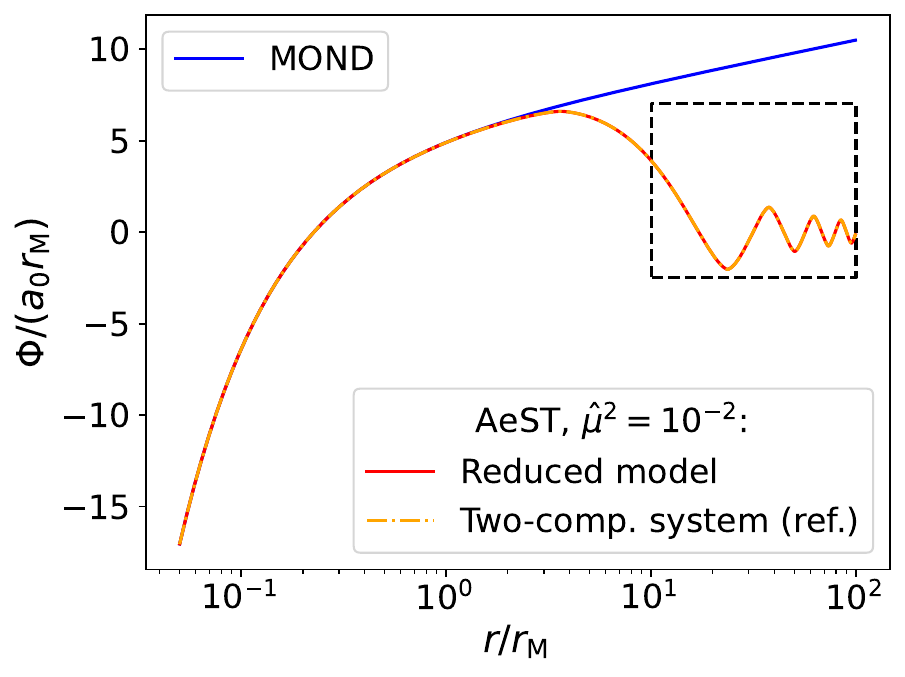}
\includegraphics[width=0.49\textwidth]{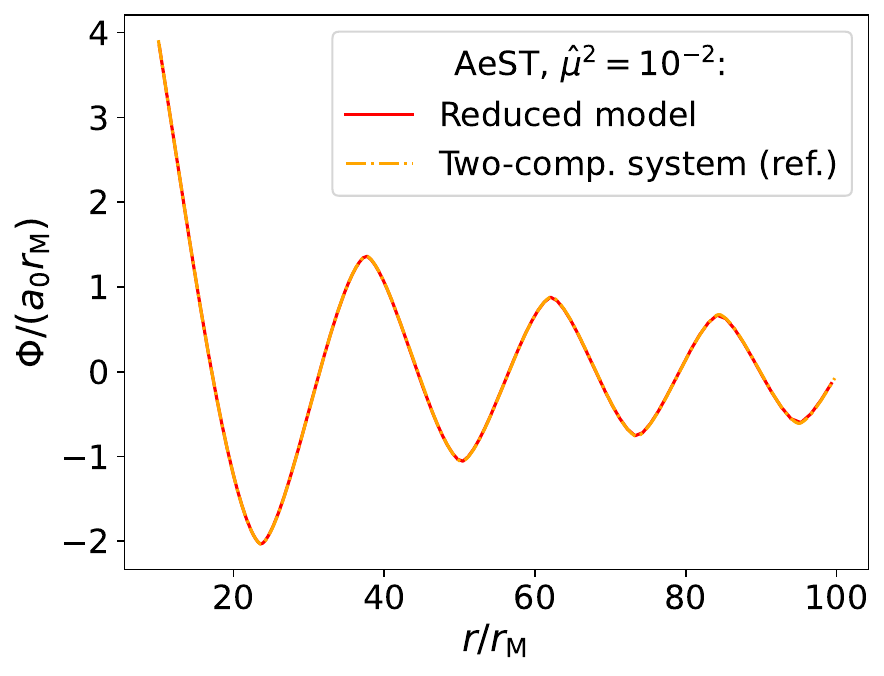}
\\
\includegraphics[width=0.49\textwidth]{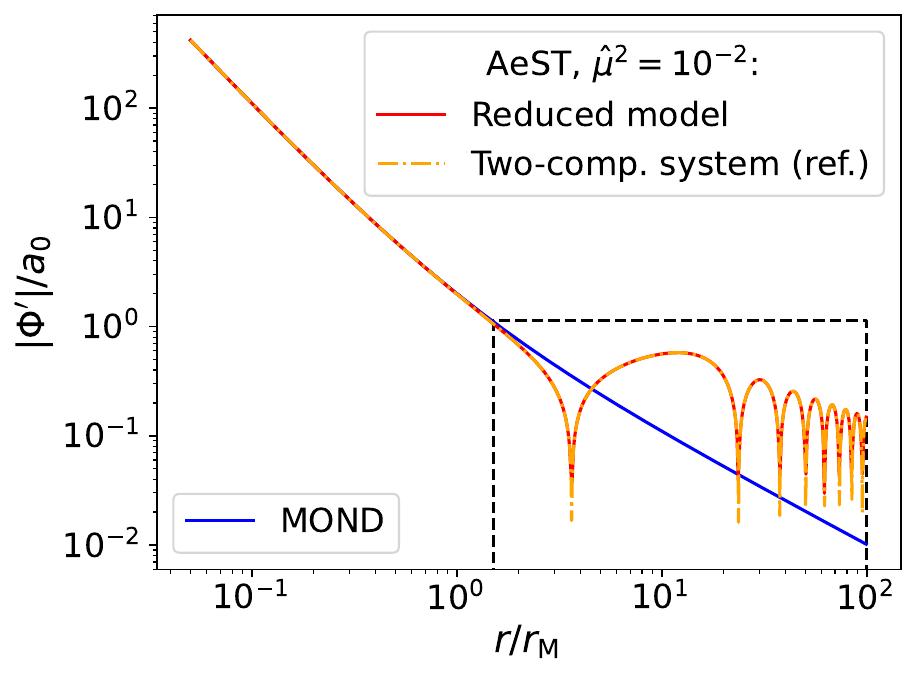}
\includegraphics[width=0.49\textwidth]{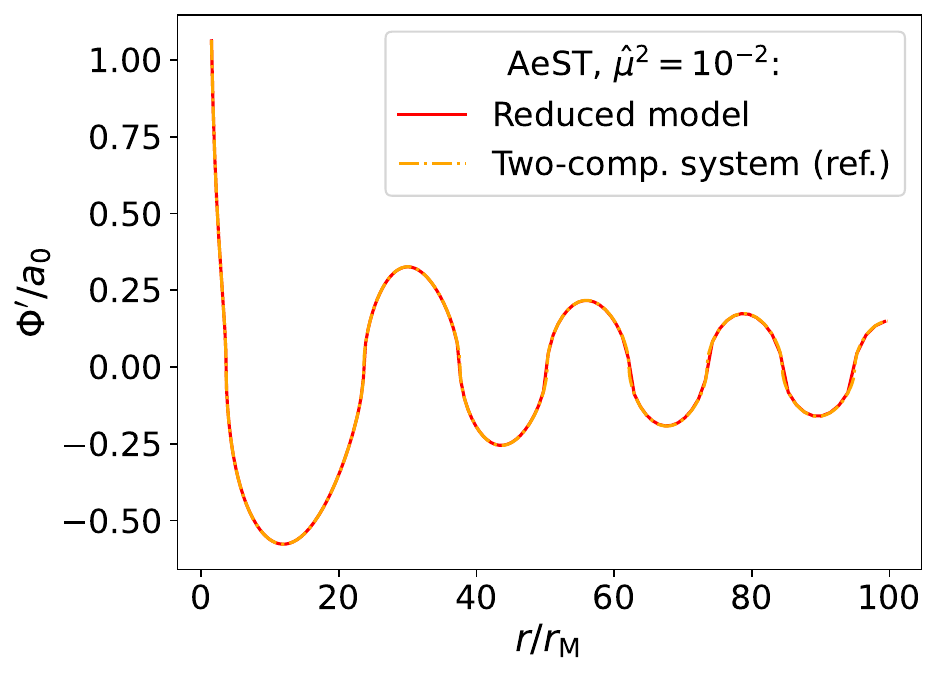}
%\subfigure{}
\caption{The momentum frame solutions (first row; left linear, right logarithmic) of $\pPhidl$ (red line) and $\pPhidl'$ (green line) of \eqref{eq_P_Phi_prime_prime_vac_dl} with clear Newtonian and MOND regimes (black dotted and dashed lines, respectively), translated into the rescaled gravitational potential $\Phidl$ (second row) and its dimensionless derivative $\Phidl'$ (third row) in AeST for $\muAeSTdl = 10^{-2}$.
The numerical solution of $\Phidl$ (red lines, second row) and $|\Phidl'|$ and $\Phidl'$ (red lines, third row) obtained from the momentum frame of the Hamiltonian system are compared with the solution of the (original) two-component system \eqref{eq:chieq2} and \eqref{eq:candiv} (orange dotted-dashed line) and pure MOND (blue lines of left column). The figures in the centre and lower right column show the oscillations in the region marked by the dashed black lines on the facing figures, but on a linear scales without absolute values.}
\label{fig:phisols}
\end{figure}

As an example we set $\muAeSTdl^2 = 10^{-2}$. We consider the gravitational potential in vacuum near a point source and 
give it Newtonian boundary conditions at $\rh_0 =  10^{-3}$ (where it is expected to be Newtonian)
 so that $\Phidl = -1/\rh_0 = -10^3$ and  $\dd\Phidl/\drh  = 1/\rh_0^2 = 10^{6}$. 
These boundary conditions for $\Phidl$ and $\Phidl'$ are then translated into boundary conditions 
for $\pPhidl$ and $\pPhidl'$ using \eqref{eq:inic1} and \eqref{eq:inic2}, or equivalently, \eqref{eq_Phi_prime_vac} and \eqref{eq_P_Phi_prime_vac}.
The numerical solution to \eqref{eq_P_Phi_prime_prime_vac_dl} of $\pPhidl$ and its translation to $\Phidl$ and $\Phidl'$ is shown in Figure~\ref{fig:phisols}.
In the Newtonian regime, $\Phi' \propto 1/r^2$ and since $\Phi' = \pPhi/\left(a_{0} r^2\right)$ this implies that $\pPhi$ is approximately constant. 
Also, $\Phi \propto -1/r$ and since $\pPhi' = -\muAeST^2 r^2 \Phi$ this implies that $\pPhi' \propto r$. In the MOND regime $\pPhi$ is also approximately constant but the growth of $\Phi$, which is only logarithmic, does not suppress the $r^2$ growth of $\pPhi'$. These two regimes are not visible in the upper left graph of Figure~\ref{fig:phisols} due to the linear axis but are visible in the upper right graph of Figure~\ref{fig:phisols} where the axes are logarithmic.

The plots of $\Phi$ and $\Phi'$ after their translation from $\pPhi$ and $\pPhi'$ using \eqref{eq_P_Phi_prime_vac_dl} and \eqref{eq_Phi_prime_vac_dl} are in the centre and lower left and right parts of Figure~\ref{fig:phisols}, respectively. They are here compared with the numerical solutions of the two-component system given by \eqref{eq:chieq2} and \eqref{eq:candiv}, and with pure MOND.
The graph of $\Phi'$ has a vertical tangent at its zeros. This is due to the vertical tangent of the function $\sign\left( \pPhidl \right) \sqrt{| \pPhidl | }$ at $\pPhidl=0$\footnote{The second derivative $\Phi''$ is therefore ill-defined which is the apparent singularity problem that the Hamiltonian formulation solves.}, as can be seen in 
\eqref{eq_Phi_prime_vac_dl}.

\begin{figure}[H]
\centering
\includegraphics[width=0.49\textwidth]{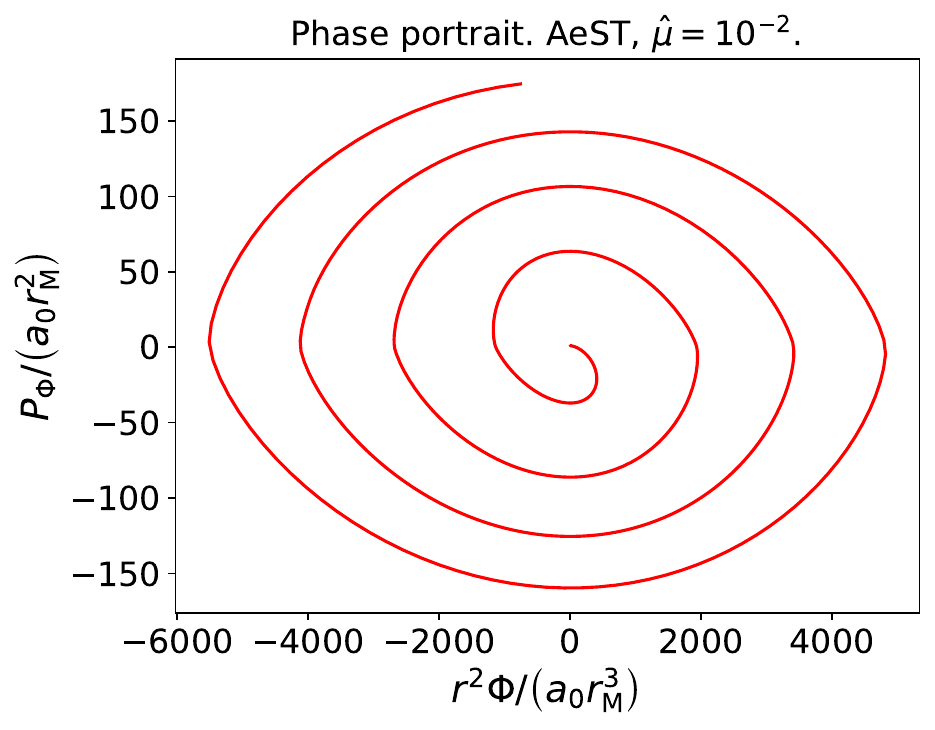}
\includegraphics[width=0.49\textwidth]{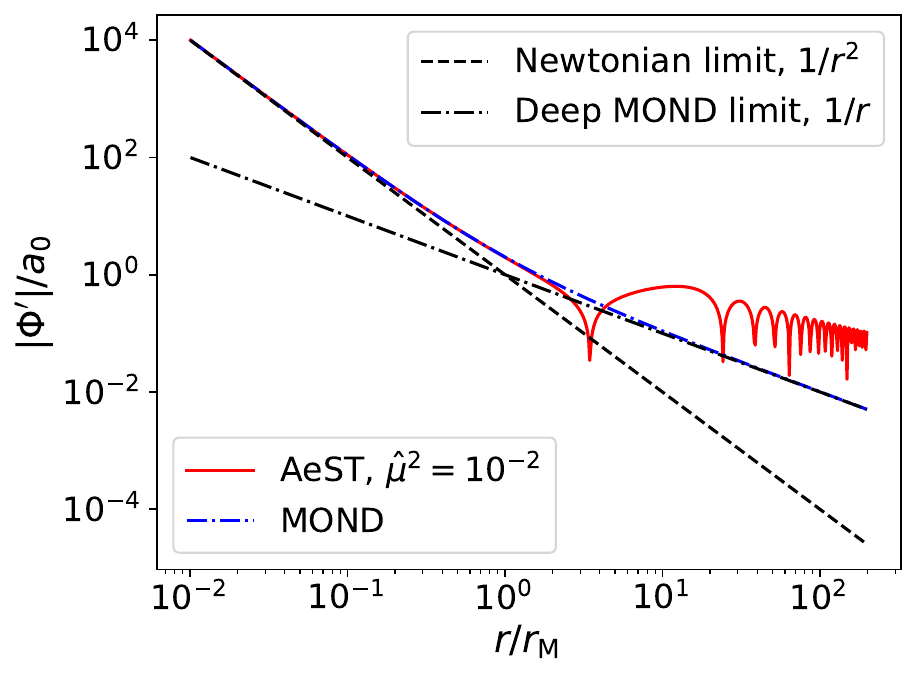}

\caption{The phase portrait $(\hat{r}^2 \hat{\Phi},\pPhidl)$ of the AeST Hamiltonian system in vacuum outside a point source with mass $M$ for (non-dimensionalised AeST weak-field mass parameter) $\muAeSTdl^2 = 10^{-2}$ (left), and the acceleration $|\hat{\Phi}'|$ (right) in AeST (red line) in vacuum outside the same source vs. radius $r/r_\textrm{M} \equiv \hat{r}$ (where $r_\textrm{M} = \sqrt{GM/a_0}$), compared to MOND (blue dashed-dotted line), with Newtonian $1/\hat{r}^2$ (dashed black line) and deep MOND $1/\hat{r}$ (dashed-dotted black line) trends indicated.}
\label{fig:ps}
\end{figure}

The phase portrait of $\Phi$ and $\pPhidl$ is shown on the left in Figure~\ref{fig:ps}, where $\Phi$ has been scaled to $r^2 \Phi$.

The numerical solution of $\Phi$ shows that, after a $1/r^{2}$ (Newtonian) decay, and a $1/r$ (MOND) decay as the MOND radius $\rh=1$ is approached, there is a new regime with oscillatory decay in $\Phi$ which begins at a scale $\rC$ determined by $\muAeST$ and initial conditions; see \eqref{eq_rC}. This regime and its dependence on
boundary conditions, interpolation function and AeST parameters $\mu$ and $\lambdas$ has been extensively studied in \cite{Verwayen:2023sds}.
A logarithmic plot of the absolute value $|\Phi'|/a_0$ versus $\rh$ reveals the trends of the decay, which numerical evidence suggests is a power-law decay of the envelope. 
This is shown on the right in Figure~\ref{fig:ps}.

\subsection{Prescribed source profile: $\Scal = \rho(r) \Phi$} 
\label{sec:pres}
The case of a (baryonic) source profile of $r$-varying density $\rho_b(r)$ is captured by setting $\Scal = \rho_b(r) \Phi$.
Using dimensionless variables as for the vacuum case,  the master equation \eqref{eq_P_Phi_prime} turns into 
\begin{align}
 \frac{\dd\pPhidl}{\drh}  =&  \rh^2 \left( - \muAeSTdl^2 \Phidl +  \rhodl_b \right)
\label{eq_P_Phi_prime_source}
%\label{eq:newp}
\end{align}
where the dimensionless density~\footnote{ This means that $\int d\rh \rh^2 \rhodl = 1$. }  $\rhodl_b = 4 \pi \sqrt{G^3 M/a_0^3}  \; \rho_b $, while \eqref{eq_Phi_prime}  remains the same as it is 
independent from $\Scal$. The second-order equation \eqref{eq_P_Phi_prime_prime_func} for $\pPhi$ is then 
\begin{align}
\frac{\dd^2\pPhidl}{\drh^2}   =&  \frac{2}{\rh}\frac{\dd\pPhidl}{\drh} 
  -  \muAeSTdl^2 \left(  \rh \sign(\pPhi) \sqrt{  |\pPhidl|  } +  \pPhidl  \right)
+   \rh^2 \frac{d\rhodl_b}{ \drh}.
\label{eq_P_Phi_prime_prime_source}
\end{align}

Curiously, in the case of a uniform density sphere of density $\rho_0$ and radius $\rstar$, since $\rho_b' = 0$ both inside and outside, the evolution is the same as the vacuum case, 
only the inside and outside solutions must be matched using \eqref{eq_P_Phi_prime_source} so as to have $\Phi$ continuous. This requires that there is a discontinuity in $\pPhi'$ 
so that
\begin{align}
\lim_{r\to \rstar^{-}} \Phi(r) = \lim_{r\to \rstar^{+}} \Phi(r) \Rightarrow \frac{4 \pi G \rho_0}{\muAeST^2} - \frac{1}{\muAeST^2 \rstar^2}    \frac{\dd\pPhi}{\dr}\bigg|_{{\rm in}} 
= - \frac{1}{\muAeST^2 \rstar^2}    \frac{\dd\pPhi}{\dr}\bigg|_{{\rm out}}  
\end{align}
and hence
\begin{align}
\Delta \pPhi' \equiv  \frac{\dd\pPhi}{\dr}\bigg|_{{\rm out}}  -  \frac{\dd\pPhi}{\dr}\bigg|_{{\rm in}} = - 4\pi G \rho_0 \rstar^2.
\end{align}

\begin{figure}[H]
\centering
\subfigure{\includegraphics[width=0.49\textwidth]{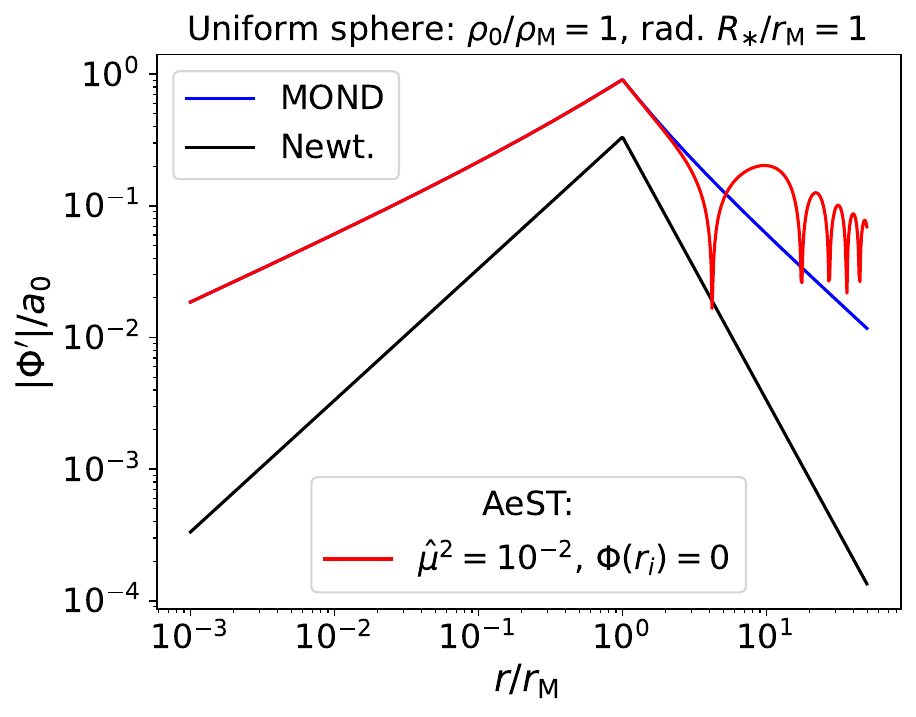}}
\subfigure{\includegraphics[width=0.49\textwidth]{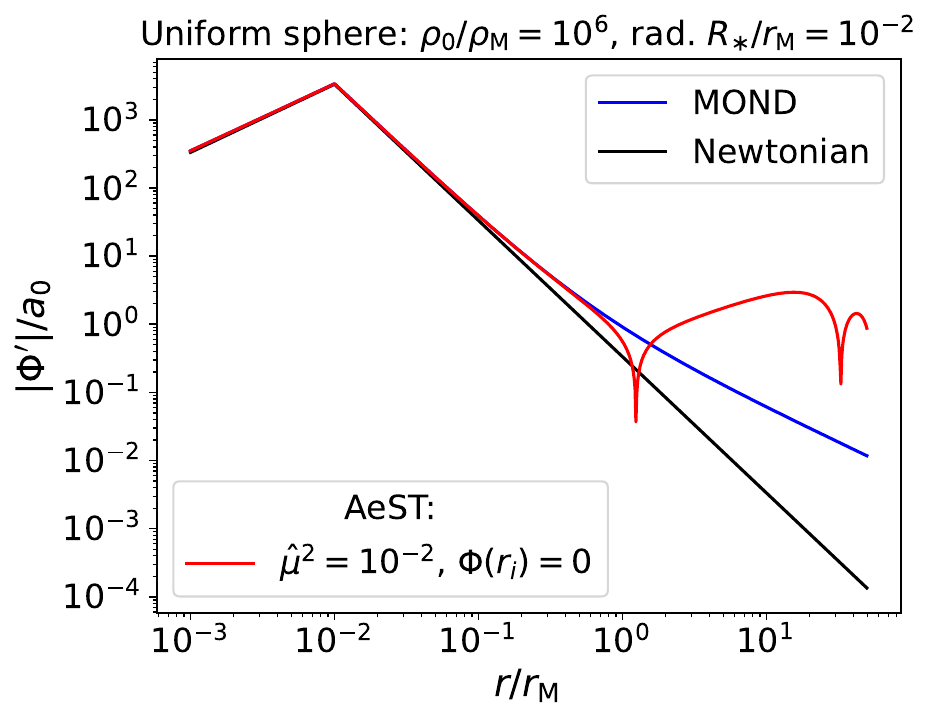}}
\caption{The acceleration ($|\Phi'|/a_0$) vs. radius for the case of the uniform density sphere in Newtonian gravity (black line), MOND (blue line) 
and AeST (for $\muAeSTdl=10^{-2}$ and $\Phi(r_i)=0$ at $r_i/r_{\mathrm{M}}=10^{-3}$, red line) for two \emph{different} radii $\rstar$ and densities $\rho_0$, low density (left figure) 
and high density (right figure), respectively, but \emph{same total mass}. For low enough densities (left figure) there is a MOND regime (black and red lines differ)
 in the inner part, not present (black and red lines coincide) with higher densities (right figure). The two cases have the same total mass and boundary conditions, 
yet the transition to oscillatory behaviour occurs earlier for the high-density case.}%, indicating a dependence, the gravitational potential, on the composition of the object.}
\label{fig:us}
\end{figure}

Although the properties of this case have been extensively treated in \cite{Verwayen:2023sds}, it is illustrative to consider, as the final self-gravitating isothermal case 
shares similar qualitative features. The case of two uniform spheres with \emph{different} densities and \emph{radii}, but \emph{same total mass}, 
are shown in Figure~\ref{fig:us} for Newtonian gravity, MOND and AeST. As more mass is enclosed the acceleration increases until all mass is enclosed. This happens abruptly for a 
uniform sphere, here at $\rh=1$ and $\rh=10^{-2}$, respectively.
If the uniform sphere is of low enough density (left figure) there is a low gradient regime throughout, coinciding with the prediction from AQUAL (since in this case $\rstar \ll \muAeST^{-1}$
and so the mass parameter does not play a role).  This is evident by the different slopes of the Newtonian (black)
 and MOND (blue) accelerations after all mass is enclosed. For high-density spheres the MOND regime is only found near $\rh \sim 1$ where the slopes of the Newtonian 
and MOND accelerations are found to differ. The acceleration in AeST follows that of MOND until a critical radius $\rC$ dependent on the gravitational potential, 
studied quantitatively in \cite{Verwayen:2023sds}, after which the acceleration becomes oscillatory. That the transition is dependent on the matter distribution
 is illustrated in Figure~\ref{fig:us} where the first node of the oscillation is found to occur earlier for the high-density case (right figure).

\subsection{Hydrostatic isothermal gas source} 
\label{sec:isoth}
\subsubsection{Hydrostatic equilibrium for isothermal profiles}
In the case of a gas in hydrostatic equilibrium with temperature $T$, the pressure gradient $\grad p$ balances the gravitational force $-\rho \grad \Phi$. 
If the gas is ideal then $p = n \kB T$ where $n$ is the number density and $\kB$ is Boltzmann's constant. Expressing
the number density in terms of the mass density as $\rho = \massgas n$, and further letting $\massgas = \mugas m_p$ where $m_p$ is 
the proton mass and $\mugas$ is the mean mass per gas particle in units of the 
proton mass~\footnote{Note that the mean mass per gas particle in units of the proton mass is less than one, which may at first seem counterintuitive, 
given that both hydrogen and helium have mass equal to or greater than that. Recall then that the gas is ionised and the electrons themselves behave as an ideal gas. Typically $\mugas = 0.6$.},
(not to be confused with AeST parameter $\muAeST$) allows us to express the pressure w.r.t. the \emph{mass} density $\rho$ as
\begin{align}
p = \frac{\rho}{\mugas m_p} k_B T.
\end{align} 
Then a first order differential equation for $\rho$ is obtained
\begin{align}
-\rho \Phi' = p' = \frac{\kB}{\mugas m_p }  \left( T  \rho' + \rho  T' \right)
 \label{eq:hydstat}.
\end{align}
For isothermal profiles that we are here considering, the temperature is constant, so that
\begin{align}
-\rho \Phi' =\frac{k_{\textrm{B} }T  }{\mugas m_p }  \rho'
\label{eq_rho_Phi}
\end{align}
leading to the solution
\begin{align}
\rho = \rho_0 e^{-\xi \left( \Phi - \Phi_0 \right) }
\end{align}
where
\begin{align}
\xi \equiv \frac{\mugas m_{p}}{ \kB T} 
\end{align}
with the boundary conditions specified at a radius $r_0$ so that $\rho(r_0) = \rho_0$ and $\Phi(r_0) = \Phi_0$.

\subsubsection{Hamiltonian formulation}
With the above expression of $\rho$ in terms of $\Phi$, we specify appropriate $\Scal$ for the case of isothermal profiles:
\begin{align}
\Scal(r,\Phi) =  -\frac{\rho_0}{\xi}  e^{-\xi \left( \Phi - \Phi_0 \right) }.
\label{Scal_iso}
\end{align}
Once again, \eqref{eq_Phi_prime} remains the same as for the previous cases while \eqref{eq_P_Phi_prime} in lieu of \eqref{Scal_iso} becomes
\begin{align}
 \pPhi' =&  r^2 \left( - \muAeSTs^2 \Phi + 4\pi \GN  \rho_0  e^{-\xi \left( \Phi - \Phi_0 \right) }  \right)
\label{eq_P_Phi_prime_iso}
\end{align}
while \eqref{eq_P_Phi_prime_prime_func} becomes
\begin{align}
 \pPhi'' =&  \frac{2}{r} \pPhi'
- r   \left(  4\pi \GN   \rho_0\xi e^{-\xi \left( \Phi - \Phi_0 \right) } + \muAeSTs^2\right) \left( \sign(\pPhi) \sqrt{ a_0 |\pPhi|  } + \frac{1}{r} \pPhi  \right).
\label{eq_P_Phi_prime_prime_func_iso}
\end{align}
In the above equation, $\Phi$ is expressed in terms of $\pPhi$ by solving \eqref{eq_P_Phi_prime_iso}, resulting in the Lambert function (product logarithm) $W$ %~\footnote{This is perhaps not surprising as $\Phi$ appears both in the exponent and linearly in the expression for $p_{\Phi}'$.}
so that
\begin{align}
\Phi =  \frac{1}{\xi} W(\Warg) - \frac{\pPhi'}{\muAeST^2 r^2} 
\label{eq:phir} 
\end{align}
where
\begin{align}
 \Warg \equiv  \frac{4 \pi G \rho_0 \xi}{\muAeST^2} e^{ \xi \left( \frac{\pPhi'}{\muAeST^2 r^2 }  + \Phi_0\right) }.
\end{align}
We now have the equation to solve \eqref{eq_P_Phi_prime_prime_func_iso} and obtain $\pPhi$ and
 the relations \eqref{eq:phiasp} and \eqref{eq:phir} to translate into $\Phi'$ and $\Phi$, respectively.

% \subsubsection{Scaling between solutions}
% A shift in $\Phi$ by $-\Phi_0$ corresponds to the same solution that would have been obtained by adding to the source a constant mass density $m^2 \Phi_0/ \left(4 \pi G \right)$ and increasing the central gas density by a factor $\exp\left(\xi \Phi_0 \right)$:
% \begin{align}
% m^2 \Phi - 4 \pi G \rho_0 \exp\left(- \xi \Phi \right) \to m^2 \Phi -  \underbrace{ \left( 4 \pi G \rho_0 \exp\left( - \xi \Phi  \right) \exp(\xi \Phi_0) + m^2 \Phi_0 \right) }_{4\pi G \rho_\mathrm{eff}}
% \end{align}

% Therefore a shift in the potential can mimic an increased gas density in the inner parts. In the outer parts the degeneracy will be broken as the volume contribution from $m^2 \Phi$ becomes significant.

\subsubsection{Isothermal profile solutions}
We introduce again dimensionless variables, albeit with some differences. First and foremost, we can no longer use the definition of the MOND radius $\rM$ from \eqref{eq_rM}, 
as it only holds in vacuum outside some source. This condition is not satisfied here, as we are dealing with a continuous density distribution which extends to 
infinity (even though in practice, as we show below, the density effectively cuts off at a finite size)
and in the Newtonian case it is well known that the total enclosed mass diverges.  As such, we need a new scale to use for creating dimensionless variables.

Our scale, denoted by $\rI$, should incorporate the constant $a_0$ so that it is relevant to the problem we are solving. It should also know about the profile (and so $c^2/a_0$, where $c$ is the speed of light 
cannot be a good choice). A good choice is the parameter $\xi^{-1}$ which has units of $c^2$ (same as $\Phi$) and involves the temperature $T$ of the profile. Let us first show that this is 
a good choice.  Starting from \eqref{eq_P_Phi_prime_prime_func_iso}, we define the dimensionless canonical momentum by $\pPhidl =  A \pPhi$ for a scale $A$ and consider the $\muAeSTs = 0$ case. Then
\eqref{eq_P_Phi_prime_iso} leads to
\begin{align}
 4\pi G  \rho_0  e^{-\xi \left( \Phi - \Phi_0 \right) }  =&  \frac{A}{\rI^3 \rh^2} \frac{\dd\pPhidl}{\drh}  
\label{rho_pPhi_MOND}
\end{align}
so that  \eqref{eq_P_Phi_prime_prime_func_iso} turns into 
\begin{align}
 \frac{\dd^2\pPhidl}{\drh^2}  =&  \frac{2}{\rh} \frac{\dd\pPhidl}{\drh}  
-    \frac{\xi A}{\rI \; \rh} \frac{\dd\pPhidl}{\drh}   \left( \sign(\pPhidl) \sqrt{ \frac{\rI^2 a_0}{A} |\pPhidl|  } + \frac{1}{\rh}   \pPhidl  \right).
\label{eq_P_Phi_prime_prime_func_iso_no_mu}
\end{align}
So the choice $A = \xi^{-1}\rI $ and 
\begin{align}
\rI \equiv \frac{1}{a_0\xi}  = \frac{k_B T}{\mugas m_p a_0}
\end{align}
is forced upon us. This choice, also prompts the definition of the dimensionless central density $\rhodl_0 \equiv  \rho_0 / \rhoI$ with $\rhoI \equiv a_0^2 \xi / (4\pi \GN) $,
the dimensionless potential $\Phidl \equiv \xi \Phi$ and the dimensionless mass parameter $\muAeSTdl \equiv \muAeSTs \rI$ as in the previous sections, 
so that the dimensionless form of \eqref{eq_P_Phi_prime_prime_func_iso}  is
%$\rhodl_0 = 4 \pi \sqrt{G^3 M/a_0^3}  \; \rho_0 $,
\begin{align}
 \frac{\dd^2\pPhidl}{\drh^2}  =&  \frac{2}{\rh} \frac{\dd\pPhidl}{\drh}  
-  \rh   \left(   \rhodl_0  e^{\Phidl_0 - \Phidl} + \muAeSTdl^2\right) \left( \sign(\pPhidl) \sqrt{  |\pPhidl|  } + \frac{1}{\rh}   \pPhidl  \right)
\label{eq_P_Phi_prime_prime_func_iso_dl}
\end{align}
where $\Phi$ is calculated from 
\begin{align}
\Phidl =   W(\Warg) - \frac{1}{\muAeSTdl^2 \rh^2} \frac{\dd\pPhidl}{\drh}
\end{align}
and where
\begin{align}
 \Warg \equiv  \frac{\rhodl_0}{\muAeSTdl^2} \; e^{ \frac{1}{\muAeSTdl^2 \rh^2 } \frac{\dd\pPhidl}{\drh}  +  \Phidl_0 }.
\end{align}
Interestingly, $\rI$ can \emph{also} be seen as the \emph{distance} over which the density decreases by one $e$-fold in the MOND regime. For the density to decrease one $e$-fold $\xi \Delta \Phi = 1$ implying $\Delta \Phi = \xi^{-1}$. In the MOND regime $\Delta \Phi / \Delta r \sim a_0$ and so $\Delta r = \Delta \Phi / a_0 = 1/\left(\xi a_0\right) \equiv \rI$.
 As we shall see below, it turns out to lead to length scales characteristic of galaxy clusters.
As a \emph{reference case} we shall take $T=5.5 \, \mathrm{keV}$ and $\mugas = 0.6$. This provides a characteristic distance scale $\rI\sim 0.24\, \mathrm{Mpc}$. 
This characteristic density $\rhoM \sim 1.9 \cdot 10^{-23} \, \mathrm{kg} \cdot \mathrm{m}^{-3} \sim 2108 \times \rho_\mathrm{crit}$ 
where $\rho_\mathrm{crit}$ is the critical cosmic energy density $\rho_\mathrm{crit} = 3H_0^2/(8 \pi \GN)$ when assuming $H_0 \sim 70 \, \mathrm{km}  \; \mathrm{s}^{-1}  \mathrm{Mpc}^{-1}$.

Before considering the full solution to \eqref{eq_P_Phi_prime_prime_func_iso_dl}, let briefly us touch upon the case $\muAeST=0$ given by \eqref{eq_P_Phi_prime_prime_func_iso_no_mu},
which in dimensionless form translates to 
\begin{align}
\frac{\dd}{\drh}\left( \frac{1}{\rh^2} \frac{\dd\pPhidl}{\drh} \right) =&  - \frac{1}{\rh^3} \frac{\dd\pPhidl}{\drh} \left[  \sign(\pPhidl) \sqrt{  |\pPhidl|  } + \frac{1}{\rh}   \pPhidl  \right]
\label{eq_P_Phi_prime_prime_func_iso_dl_no_mu}
\end{align}
where we have also set $\muAeSTs=0$ into \eqref{eq_P_Phi_prime_iso} to obtain $\dd\pPhidl/\drh =  \rhodl_0  \rh^2 e^{\Phidl_0 - \Phidl}$. This equation has no free parameters! 

Neglecting the first term on the RHS in \eqref{eq_P_Phi_prime_prime_func_iso_dl_no_mu} (the term containing $\sqrt{|\pPhidl|}$) amounts to neglecting the MOND regime, that is, it corresponds to the
Newtonian isothermal sphere. The general analytical solution is not known, however, the particular solution $\pPhidl = 2 \rh$ corresponds to the singular isothermal sphere~\cite{BinneyTremaineBook}. 
Neglecting the Newtonian part and focussing on the MOND contribution coming from the term containing the  $\sqrt{|\pPhidl|}$ leads to
\begin{align}
\frac{\dd}{\drh}\left( \frac{1}{\rh^2} \frac{\dd\pPhidl}{\drh} \right) =&  - \frac{1}{\rh^3} \frac{\dd\pPhidl}{\drh}   \sign(\pPhidl) \sqrt{  |\pPhidl|  }.
\label{eq_P_Phi_prime_prime_func_iso_dl_no_mu_MOND}
\end{align}
The general solution in this case contains implicit functions, however, a particular solution is found to be  (this is \eqref{Core_ISO_1})
\begin{align}
\pPhidl = \frac{81}{4} \frac{ \rh^3/\rh_0^3 }{\left(\sqrt{ \rh^3/\rh_0^3} + 1\right)^2 } \qquad \forall \quad \rh>0
\end{align}
which corresponds to a cored density profile controlled by the parameter $\rh_0$, given by
\begin{align}
\rhodl = \frac{243 a_0^2 \xi}{16\pi \GN \rh_0^3 \left[1 + (\rh/\rh_0)^{3/2}\right]^3}
\end{align}
and potential as
\begin{align}
\Phi = \Phi_0 +\frac{1}{\xi} \ln \frac{4\rhodl_0\rh_0^3}{243} + \frac{3}{\xi}\ln \left[1 + (\rh/\rh_0)^{3/2}\right].
\end{align}
The details of the derivation of the pure MOND isothermal profile are found in Appendix~\ref{Pure_MOND_iso}.

Interestingly, $\pPhi$ is nothing but  the mass enclosed within radius $r$. That is, $M = 4\pi \int_0^r \dr r^2 \rho(r) $ leads to
$\GN M =    \int_0^r \dr \frac{\dd\pPhi}{\dr} = \pPhi(r) - \pPhi(0)$. Hence, taking $\rh \rightarrow \infty$, the total mass enclosed is
\begin{align}
  M=    \frac{81}{4 \GN a_0\xi^2  }  =   \frac{81 \rI^2 a_0}{4 \GN}.
\end{align}
Hence we find that
\begin{align}
  \rI = \frac{2}{9} \sqrt{ \frac{\GN M}{a_0}} = \frac{2}{9} \rM
\end{align}
is the MOND radius in disguise! Interestingly though, the MOND radius $\rM$ was defined in vacuum outside a source of mass $M$, while here, we have a continuous source and $M$ is only found after 
integrating to $r=\infty$. Nevertheless, the relevant scale remains the same.

We now describe the full numerical solution found by integrating \eqref{eq_P_Phi_prime_prime_func_iso_dl}.  
The numerical procedure is described in Appendix~\ref{sec:numsol} and  for ensuring robustness of results, we compare to the two-component system and demonstrate excellent agreement; see Figure~\ref{fig:checks} of Appendix~\ref{sec:redone}.
\begin{figure}[H]
\centering
\includegraphics[width=0.49\textwidth]{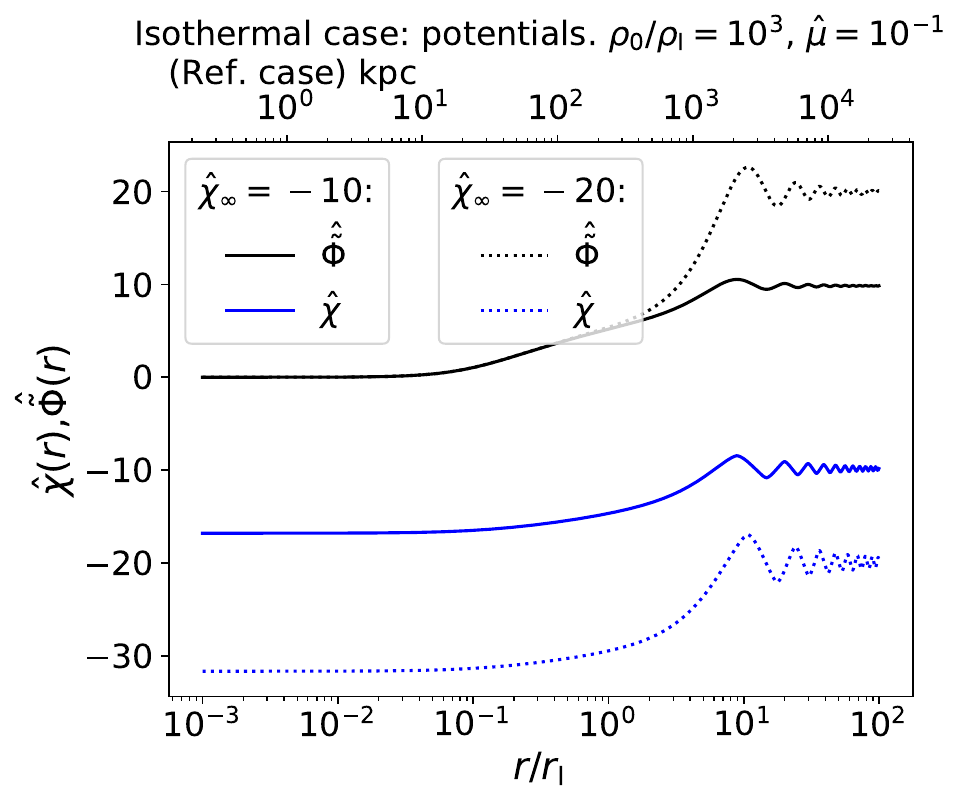}
\includegraphics[width=0.49\textwidth]{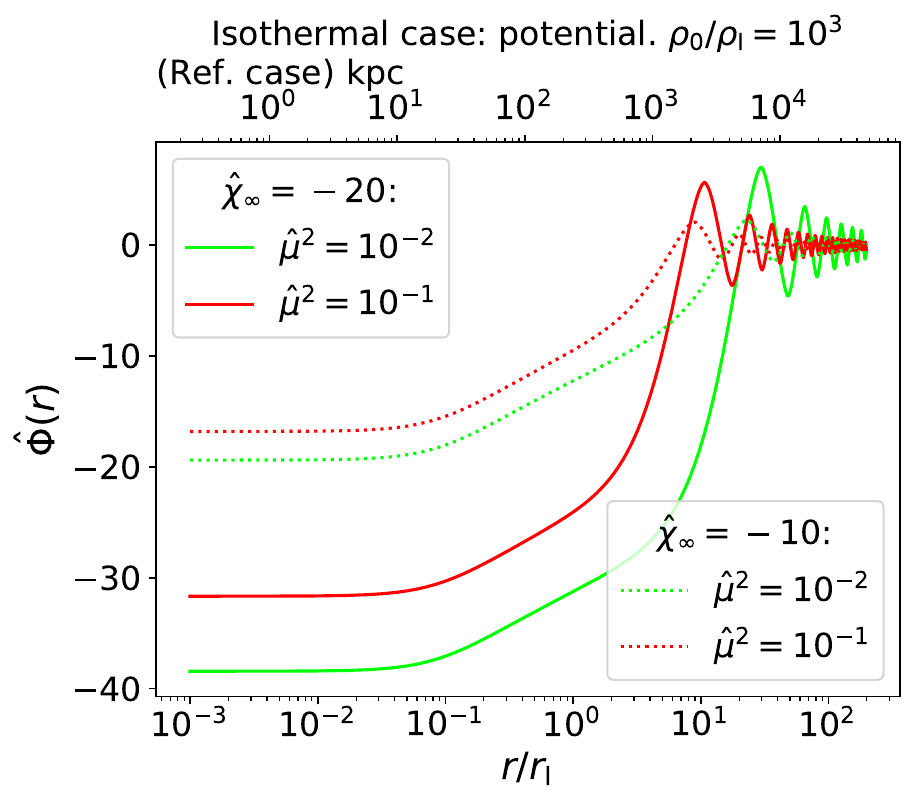}
\caption{{\bf Left panel:} The AeST potentials $\hat{\chi}(r)$ (blue lines) and $\hat{\Phit}(r)$ (black lines) for the case of the hydrostatic isothermal gas with central density $\rhodl_0= \rho_0/\rhoI = 10^{3}$
for $\muAeSTdl=10^{-1}$ and two choices of the asymptotic value of $\hat{\chi}(r)$: $\hat{\chi}_{\infty}=-20$ (dotted lines) and $\hat{\chi}_\infty=-10$ (full lines). 
Interestingly, the asymptotic values of the two fields are negatives of one another $\hat{\chi}_{\infty}=-\hat{\Phit}_{\infty}$ so that $\hat{\Phi}\sim 0$.
{\bf Right panel:} The gravitational potential $\hat{\Phi}(r)\equiv \hat{\chi}(r) + \hat{\Phit}(r)$ for two different choices of the AeST 
mass parameter, $\muAeSTdl= 10^{-2}$ (green line) and $\muAeSTdl = 10^{-1}$ (red line), and two different choices of the asymptotic values $\hat{\chi}_{\infty} = -20$ (solid) and $\hat{\chi}_{\infty}=-10$ (dotted). 
The gravitational potential $\hat{\Phi}(r)$ decays while oscillating about zero. The top axes indicate the physical length for the reference case of a $T=5.5 \, \mathrm{keV}$ ideal gas.}
\label{fig:potentials}
\end{figure}

In order to impose boundary conditions, but also to compute $\Phi$ which is needed in \eqref{eq_P_Phi_prime_prime_func_iso_dl}, we use the algebraic relations between $\Phidl$ and $\dd\pPhidl/\drh$ as well as 
between  $\dd\Phidl/\drh$ and  $\pPhidl$.  These can be found from the results of the previous sections and are \eqref{relation_Phi_prime_P} and \eqref{relation_Phi_Pi}  respectively. 
Having $\Phidl$ and $\dd\Phidl/\drh$, we can then reconstruct $\dd\Phitdl/\drh$ and $\dd\chidl/\drh$ through the interpolation function and further integrate them to obtain $\Phitdl$ and $\chidl$,
and remember that $\Phidl = \Phitdl + \chidl$. Here the ``hats'' denote dimensionless quantities by rescaling with $\xi$ as in the case of $\Phidl$. 

We choose boundary conditions such that  $\Phitdl(\rh_{\ini}) = 0$  and  $\chidl(\rh_{\ini}) = \chidl_{\ini}$ so that $\Phidl_{\ini} = \Phitdl + \chidl = \chidl_{\ini}$. Rather than 
$\chidl_{\ini}$ it is sometimes desirable to refer to the final boundary condition as $r\rightarrow \infty$, that is, $\chidl_{\infty}$, as we use in the figures. The boundary condition for $\dd\Phidl/\drh$ as well
 as the connection to the boundary conditions for $\pPhidl$ and $\dd\pPhidl/\drh$ are found in appendix  \ref{sec:numsol}.
%For negative boundary conditions of $\Phi(r_0)<0$, $\Phi \to 0$ as $r \to \infty$ so $\chi_{\infty} = - \Phit_\infty$ and so $\chi_\infty$ can be used to describe the boundary condition. For positive initial conditions $\Phi(r_0)>0$, $\Phi \to \Phi_{\infty} \neq 0$, and so then $\Phi_{\infty}$ is more appropriate to use to describe the boundary condition.
We set $\rhodl_0 = 10^{3}$ which ensures a dense enough state that the high acceleration (Newtonian regime) is valid within the inner part of the isothermal sphere.

The potentials $\hat{\Phit}(r)$, $\hat{\chi}(r)$ and $\hat{\Phi}(r)$ for two cases of $\muAeSTdl$ and two different boundary conditions $\chidl_{\infty}$  are shown in Figure~\ref{fig:potentials}.
Just like the vacuum and prescribed source case of the previous sections, beyond a certain radius the potentials oscillate with $r$. However, here it may happen that
the oscillations are around a non-zero asymptotic value, and this depends on the boundary condition.

\begin{figure}[H]
\centering
\includegraphics[width=0.52\textwidth]{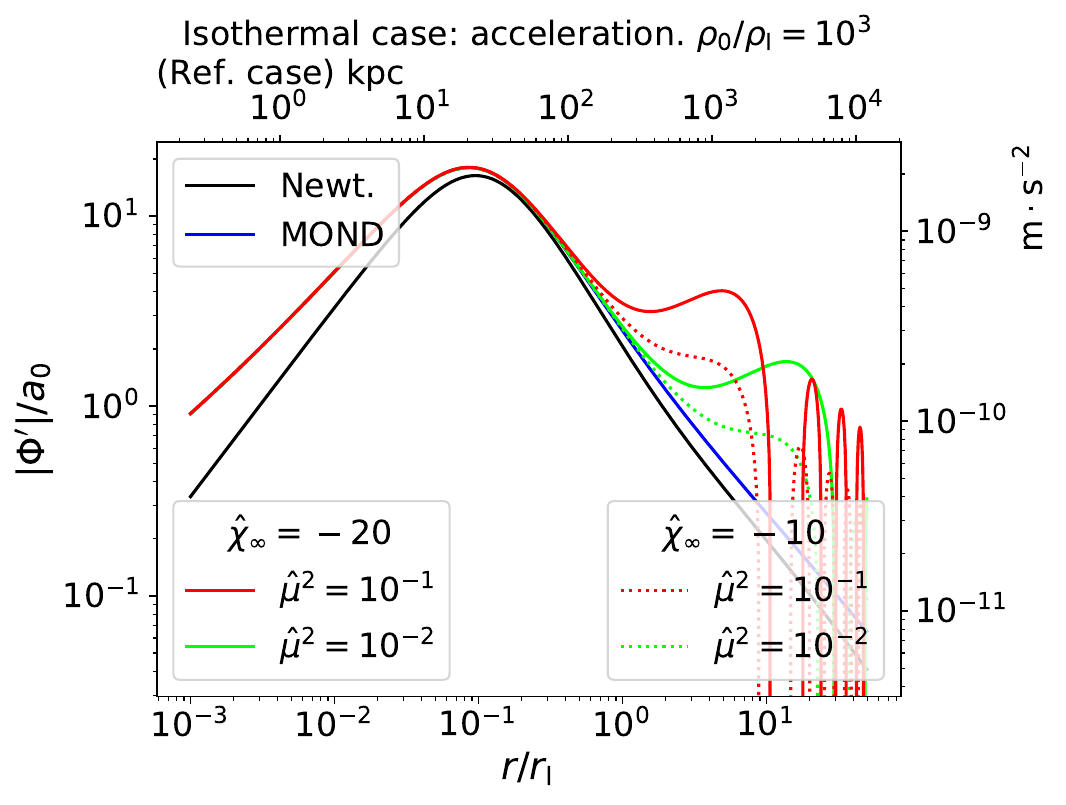}
\includegraphics[width=0.46\textwidth]{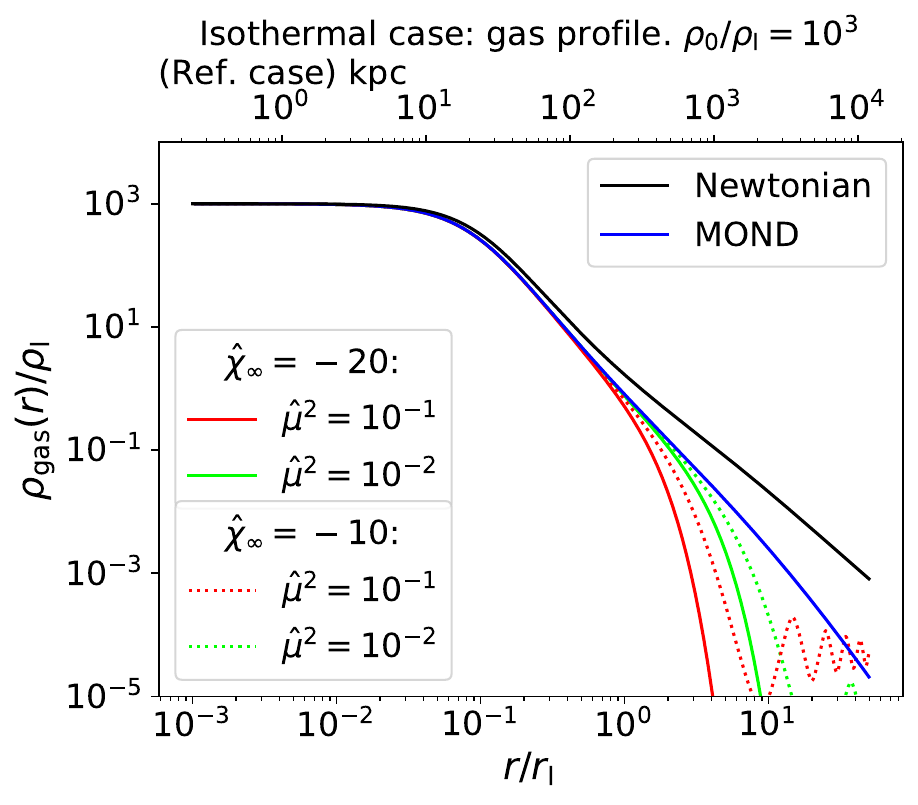}
\caption{{\bf Left panel:} Acceleration $|\Phi'|/a_0$ vs. radius $r/\rI$ for the hydrostatic isothermal gas with initial density $\rho_0/\rhoI=10^{3}$ in Newtonian gravity (black line), MOND (blue line) and
the same  AeST models as in Figure~\ref{fig:potentials}: $\muAeSTdl=10^{-1}$ in red  and $\muAeSTdl=10^{-2}$  in green with solid corresponding to  $\chi_{\infty} = -20$  and dotted corresponding to $\chi_{\infty} = -10$.
{\bf Right panel:} Gas density profiles for the hydrostatic isothermal gas in AeST, compared to MOND and Newtonian gravity for the same models as in the left panel.
The density profiles of AeST become more compressed. For the model case  $\{\muAeSTdl=10^{-1},\chi_{\infty} = -10\}$ one sees the oscillations of the potential imprinted on the density profile.
 The top axes in both panels indicate the physical length for the reference case of a $T=5.5 \, \mathrm{keV}$ gas.
}
\label{fig:acc_vs_r_wpx}
\end{figure}
% Here, it is seen how the asymptotic values of $\chi$ and $\Phit$, $\chi_{\infty}$ and $\Phit_{\infty}$, are the negatives of each other.
On the left panel of Figure~\ref{fig:acc_vs_r_wpx} we display the potential gradient $|\Phi'|/a_0$ which corresponds to the acceleration felt by a gas particle and on the right panel of
Figure~\ref{fig:acc_vs_r_wpx}  we show the gas density profile.
It is seen that AeST for shifted potentials $\chi_{\infty}<0$, provides a stronger acceleration than MOND. A greater shift in the potential produces a greater deviation from MOND throughout the extended source. However, beyond a certain radius, dependent on $\muAeSTdl$ and $\chi_{\infty}$, the acceleration goes below the MOND expectation, and then oscillates.
It is seen that the gas profiles are significantly more compressed in AeST compared to either Newtonian gravity or MOND. At large radii and small gas densities the oscillations from the potential $\Phi$
are  imprinted onto the gas density, effectively trapping the gas at the minima and rarefying it at the maxima of the potential.

\subsubsection{Effective phantom mass}
\begin{figure}[H]
\centering 
\includegraphics[width=0.49\textwidth]{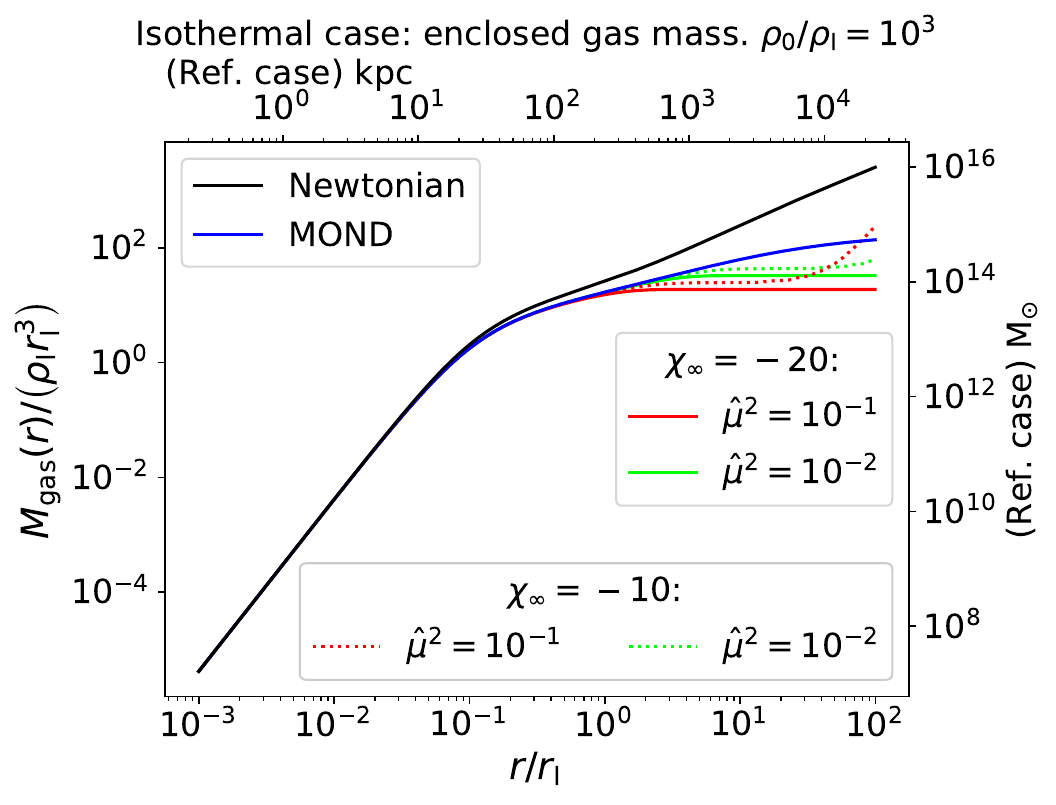}
\includegraphics[width=0.49\textwidth]{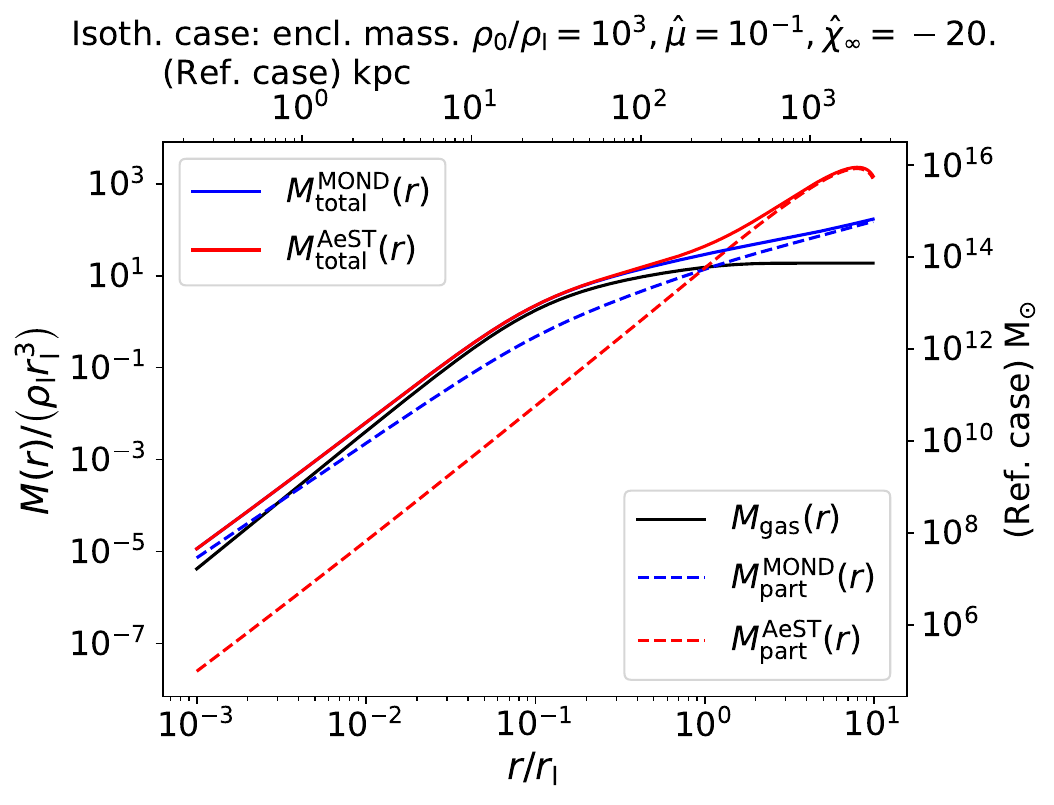}
\caption{{\bf Left panel:} The mass profile for the gas $M_{\mathrm{gas}}(r)$ for the same models as in Figure~\ref{fig:acc_vs_r_wpx}: AeST (red and green lines), MOND (blue line) and Newtonian gravity (black line). 
{\bf Right panel}: The gravitational force of AeST interpreted as an effective mass in Newtonian gravity due to the MOND part of the theory $M^\mathrm{MOND}_{\mathrm{part}}(r)$ (dashed blue line) and the AeST-specific part $M^{\mathrm{AeST}}_{\mathrm{part}}(r)$ (red dashed line) in addition to the gas mass $M_{\mathrm{gas}}(r)$ (black line), giving the total MOND mass $M^{\mathrm{MOND}}_{\mathrm{total}}(r)$ (gas and phantom mass, full blue line) 
and the total enclosed AeST mass $M^{\mathrm{AeST}}_{\mathrm{total}}(r)$ (gas, phantom and $\muAeST$-mass, full red line), respectively. 
The main AeST contribution peaks in the outer part. The top axes indicate the physical length in $\mathrm{kpc}$ and the right axes the mass in units of the solar mass $M_{\odot}$ for the reference case 
of a $T=5.5 \, \mathrm{keV}$ gas.} 
\label{fig:masp}
\end{figure}

It is instructive to consider the would-be mass contributed by MOND and AeST, respectively. In the MOND context, this is referred to as the \emph{phantom mass}. The total apparent mass in the system is defined as
the effective source of $\Phi$, if it obeyed the Poisson equation. That is, one determines $\Phi$ for a given model (e.g. MOND or AeST), then constructs the $r$-dependent effective total mass as
\begin{align}
M_{\mathrm{total}}(r) \equiv -r^2 \Phi'/G.
\label{M_total}
\end{align}
In order to be able to compare AeST to MOND, we first run an AeST model and compute the corresponding gas density $\rho_{b}(r)$ and potential $\Phi^{(\mathrm{AeST})}(r)$. Given this  $\rho_{b}(r)$,
we use it as a prescribed source in the $\muAeST = 0$ (i.e. MOND) equations of the previous section and determine the equivalent MOND potential $\Phi^{( \mathrm{ MOND} )}(r)$. From these two potentials,
we determine $M^{(\mathrm{AeST} )}_{\mathrm{total}}$  and $M^{(\mathrm{ MOND})}_{\mathrm{total}}$  according to \eqref{M_total}.
The enclosed MOND $\chi$-specific part, the phantom mass, is then computed from 
\begin{align}
M^{\mathrm{MOND}}_{\mathrm{part}}(r) \equiv M^{\mathrm{MOND}}_{\mathrm{total}}(r) - M_{\mathrm{gas}}(r)
\label{M_part}
\end{align}
while the purely  AeST part, that is, the contribution from $\muAeST^2$, excluding the pure MOND contribution and the gas, is  computed as
\begin{align}
M^{\mathrm{AeST}}_{\mathrm{part}}(r) \equiv M^{\mathrm{AeST}}_{\mathrm{total}}(r) - M^{\mathrm{MOND}}_{\mathrm{total}}(r).
\end{align}
The resulting enclosed mass profiles are shown in Figure~\ref{fig:masp}.

It is seen that the AeST contribution to the apparent mass is more significant in the outskirts, growing steadily until the effects of $\muAeST^2 \Phi$ become significant. 
%The AeST contribution to the enclosed mass grows almost as $R^3$, i.e., $M^{\mathrm{AeST} }_{\mathrm{part}}(R) \propto R^3$. This is seen in Figure~\ref{fig:trend}.
%\begin{figure}[H]
%\centering 
%\includegraphics[width=0.6\textwidth]{mass-decomposition_wtrend.pdf}
%\caption{The trend of the AeST contribution to the \emph{effective} enclosed mass. The AeST contribution $M^{\mathrm{AeST} }_{\mathrm{part} }$ (dashed red line) to the \emph{effective} enclosed mass $M^{\mathrm{AeST} }_{\mathrm{total} }$ (full red line) is seen to grow almost as $r^3$ (dashed black line). The $r^2$ trend (dashed dotted line) is indicated for reference. The enclosed baryonic mass $M_{\mathrm{gas}}(r)$ (full black line) is also shown. The top axes indicate the physical length in $\mathrm{kpc}$ and the right axes the mass in units of the solar mass $M_{\odot}$ for the reference case of a $T=5.5 \, \mathrm{keV}$ gas.} %, equivalently, by multiplication by temperature the pressure profiles, the density profile extended to large radii illustrating the gas trapped in the potential oscillations (upper right), total enclosed gas mass (upper left) and mass decomposition (lower right).}
%\label{fig:trend}
%\end{figure}
%This implies that the AeST contribution is as a constant mass density throughout the source except in the outskirts.

\subsection{The Radial Acceleration Relation of Aether-Scalar-Tensor theory} \label{sec:rar}

The Radial Acceleration Relation (RAR) is the relation between the actual acceleration, and the acceleration that would be inferred from just the given matter distribution, in principle visible, in Newtonian gravity.
In AeST, there is no \emph{universal} RAR, as the mass term $\muAeST^2 \Phi$ introduces effects that depend on the matter distribution and the boundary condition $\chi_{\infty}$.

\begin{figure}[H]
\centering
\includegraphics[width=0.49\textwidth]{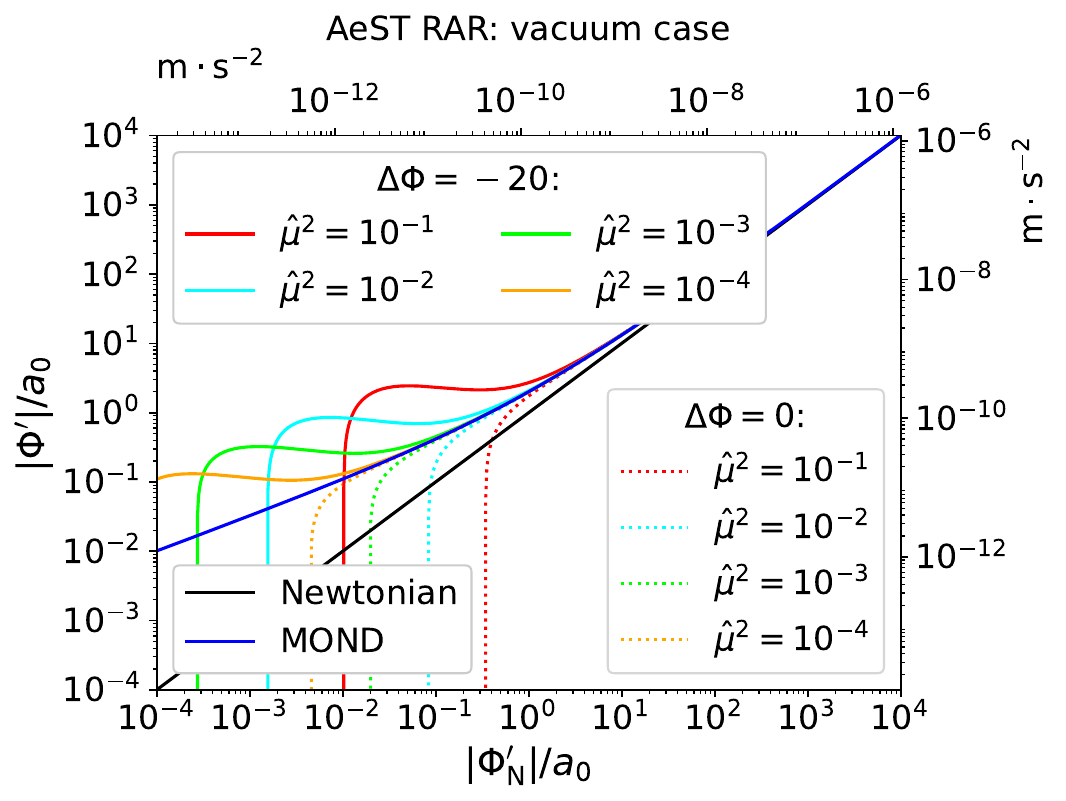}
\includegraphics[width=0.49\textwidth]{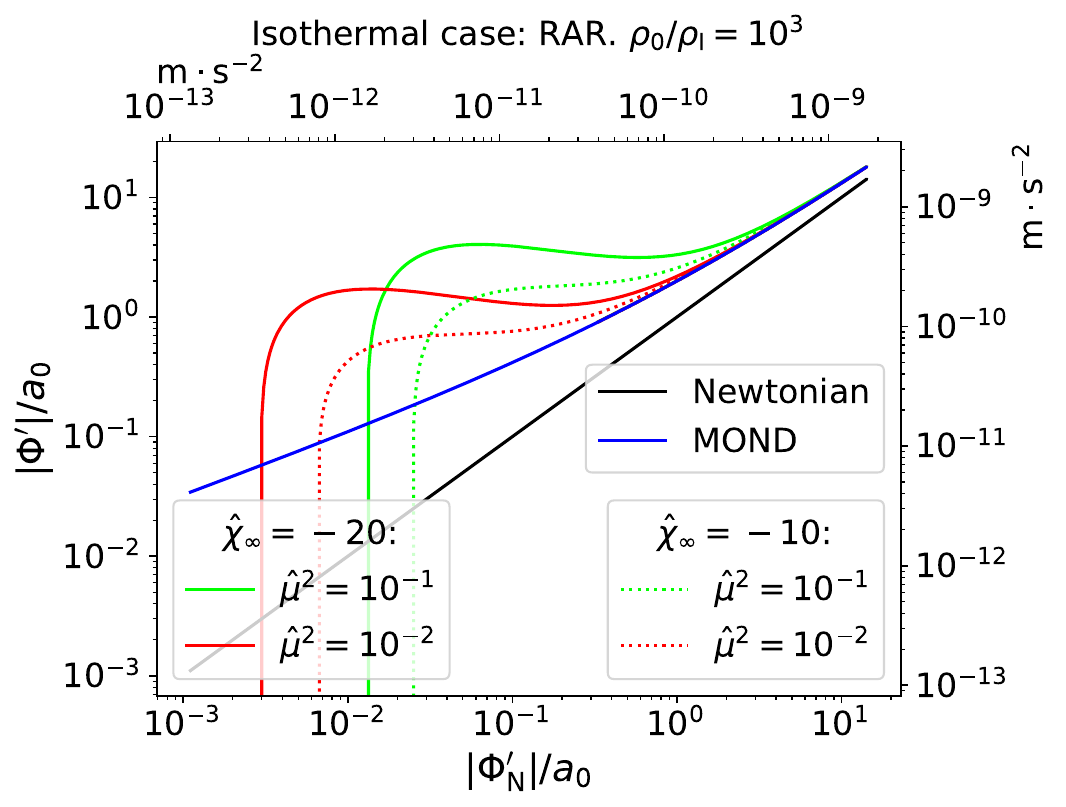}
\caption{The AeST Radial Acceleration Relation acceleration for the vacuum case (left) and isothermal sphere (right) with $\rho_{0}/\rhoI=10^{3}$.
The AeST RAR depends on the mass of the system and the shift of the potential $\Delta\Phi$ (full vs dotted lines). The RAR is shown for Newtonian gravity (trivial, black line with slope 1),
 MOND (blue line), and various choices of $\muAeSTdl$ and boundary condition for the potential ($\Delta \Phi$ in the vacuum case and $\chi_{\infty}$ in the isothermal case).
In both cases the RAR deviates from MOND in the low acceleration regime and may display a peak followed by a sharp descent below it.
 Smaller values of $\muAeSTdl$ push the peak to smaller accelerations and the peak may disappear altogether, with only the sharp descent remaining, for some boundary conditions.
}
\label{fig:rar}
\end{figure}

The RAR for the vacuum solutions is shown on the left panel of in Figure~\ref{fig:rar}.
A peak goes above the MOND expectation in the case of AeST for  potentials shifted by the boundary condition for $\Phi$. The size and position of this peak depend on both the shift in the potential $\Delta \Phi$ and the dimensionless mass parameter $\muAeSTdl$. As the square mass parameter $\muAeSTdl$ is proportional to the mass of the source through the dependence of $\muAeSTdl^2 \equiv \muAeST^2\rM^2$ on $\rM \propto M$ it 
implies that the transition to AeST behaviour is \emph{mass-dependent}. The peak is made larger by a more negative shift of the potential. 
Inevitably, beyond a certain range the AeST RAR goes below the MOND RAR. In the context of MOND this has the appearance of a negative mass density.

A similar RAR is present for isothermal gaseous spheres, which can function as a simplified model of a galaxy cluster, as shown on the right panel of Figure~\ref{fig:rar} for  $\rho_{0}/\rhoI=10^{3}$ 
and two different choices of $\muAeSTdl$ and two different boundary conditions parametrised by $\hat{\chi}_{\infty}$. The quantitative features and behaviour with respect to changing $\muAeSTdl$ 
and $\hat{\chi}_{\infty}$ are similar as in the vacuum case.

Interesting behaviour for the isothermal case emerges when the density is so low, that the solution is in the low acceleration regime throughout. In that case, the RAR displays 
 bifurcation such that there are two observed accelerations $\Phi'$ for a given Newtonian acceleration $\Phi'_\mathrm{N}$. We discuss this in Appendix~\ref{sec:rarapp}.

\section{Conclusion}
We studied the weak-field spherically symmetric static solutions in AeST theory, extending previous work done in~\cite{Verwayen:2023sds} and~\cite{Mistele:2023fwd}.
We showed that if boundary conditions allow curls to be neglected, such as the case of spherical symmetry, the two-component version of the AeST  equations in the
weak-field spherically symmetric static limit can be mapped onto a one-component equation for the metric potential $\Phi$ resembling the standard AQUAL equation plus a mass 
term which is new and coming from AeST theory.

The resulting one-component equation for $\Phi$ seems to lead to singularities in the evolution when its gradients $|\grad\Phi|$ go through zero, which happens in AeST because the mass term $\muAeST^2 \Phi$
induces oscillations in $\Phi$ as was already reported in~\cite{Verwayen:2023sds} and~\cite{Mistele:2023fwd}. We showed that these are only apparent singularities and nothing physically bad happens at those points. 
By creating an equivalent Hamiltonian system, and evolving the canonical momentum $\pPhi$ rather than $\Phi$, the resulting equation does not have any points where a term may become singular. 

We solved the two-component field-based and one-component Hamiltonian system numerically in three cases: in vacuum outside a spherically symmetric source, within a prescribed source and
for an isothermal sphere in hydrostatic equilibrium which serves as  a simplified model of a galaxy cluster.
The vacuum solutions have a Newtonian regime with force scaling as $1/r^2$, an intermediate MOND regime with force scaling as $1/r$ and an oscillatory phase with an envelope that decays as a power law.
For isothermal spheres, the gas profiles in AeST were found to be more compressed than both the Newtonian case and in the case of MOND, signifying a stronger gravitational force than in MOND, or equivalently, 
more apparent dark matter.  In all cases, we found excellent agreement between the two methods. 

We constructed the Radial Acceleration Relation (RAR) for the isothermal hydrostatic source. For shifted gravitational potentials we found that there is a peak in the RAR, an enhancement of the acceleration in a range determined by the shift, the weak-field mass parameter of AeST and the mass of the system. The peak in the AeST RAR compared to MOND is followed by a deficit with the MOND expectation, interpreted as a negative phantom mass in MOND. %The enhancement and, though to a lesser degree, the deficit are found in the (observational) galaxy cluster RAR.

Our analysis of the isothermal sphere indicates that AeST possesses the qualitative features to address the problem of galaxy clusters in MOND.
 The (observational) galaxy cluster RAR~\cite{Tian:2020qjd,Freundlich:2021jie,Li:2023zua} displayes a peak above the MOND expectation, with hints of deficit below the MOND expectation (negative phantom mass) in the outskirts. 
We found such behaviour to happen in AeST, however, it remains to be seen whether this effect can be corroborated with real data after using realistic galaxy cluster models in AeST.
A quantitative analysis going beyond the isothermal case, and the presence of multiple components needed for the analysis of a realistic galaxy cluster is left for future work.

\acknowledgments

We thank Benoît Famaey, Tom Złośnik and William Barker for useful discussions.
AD, and CS in part, were supported by the European Structural and Investment Funds and the Czech Ministry of Education, Youth and Sports:
Project CoGraDS - CZ.02.1.01/0.0/0.0/15003/0000437.
AD was from 2023/03/01 supported by the European Regional Development Fund and the Czech Ministry of Education, Youth and Sports:
Project MSCA Fellowship CZ FZU I - CZ.02.01.01/00/22\_010/0002906.

\bibliographystyle{JHEP}
\bibliography{refs}

\appendix

\section{Interpolation functions and screening}
\label{sec:screening}
In this appendix we present two further interpolation functions: one for which $\lambdas\rightarrow \infty$ results in the totally screened function \eqref{eq:simple_interp}
 and one corresponding to the ``simple'' MOND interpolation function.

\subsection{Generalization of our totally screened interpolation function \eqref{eq:simple_interp}}
Consider the function 
\begin{align}
\Jcal =&
 \lambdas \bigg\{\Ycal - 2 a_0 (1+\lambdas) \sqrt{\Ycal} 
+2 (1+\lambdas)^2a_0^2 \ln\left[1  +  \frac{\sqrt{\Ycal}}{ (1+\lambdas)a_0}\right]\bigg\} 
\end{align}
which upon further inspection has the correct Newtonian and MOND limits given by \eqref{def_lambdas} and \eqref{Jcal_MOND_limit} respectively.
Taking derivatives we find that
\begin{align}
\Jcal_{\Ycal} =& \frac{\lambdas \sqrt{\Ycal}}{a_0(1+\lambdas) + \sqrt{\Ycal}}.
\end{align}
Following the procedure in section \ref{sec:oneeq} this leads to the interpolation function
\begin{align}
\interp(x) = \frac{1+\lambdas + (1+2\lambdas) x -\sqrt{ (\lambdas +1)^2  +2 (\lambdas +1) (2 \lambdas +1) x + x^2} }{2 \lambdas  x}
\end{align}
which has the right MOND ($\interp \rightarrow x$) and Newtonian ($\interp \rightarrow 1$) limits. Now taking the $\lambdas\rightarrow \infty$ limit we find that
\begin{align}
\interp(x) \rightarrow \frac{1 + 2 x - \sqrt{1  +4  x} }{2 x} = \frac{-1+\sqrt{1+4x}}{1+\sqrt{1+4x}}
\end{align}
which is \eqref{eq:simple_interp}. At the same limit, we have that $\Jcal_{\Ycal}\rightarrow   \sqrt{\Ycal}/a_0$ (which when
integrated would correspond to \eqref{eq:simpcho}), although $\Jcal$ diverges.

\begin{figure}[H]
\centering
%\subfigure{}
\includegraphics[width=0.49\textwidth]{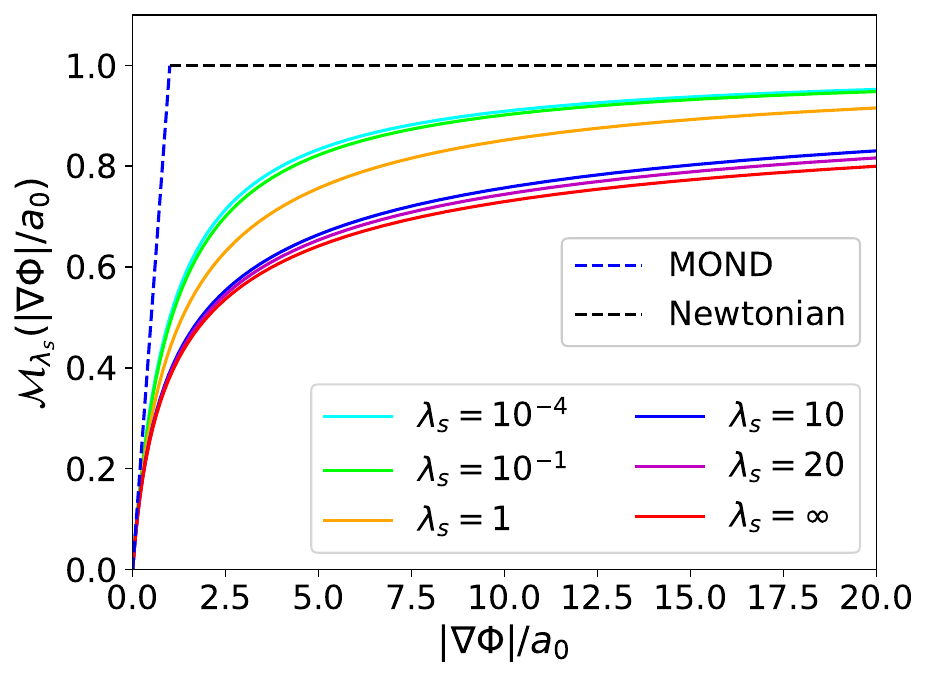}
\includegraphics[width=0.49\textwidth]{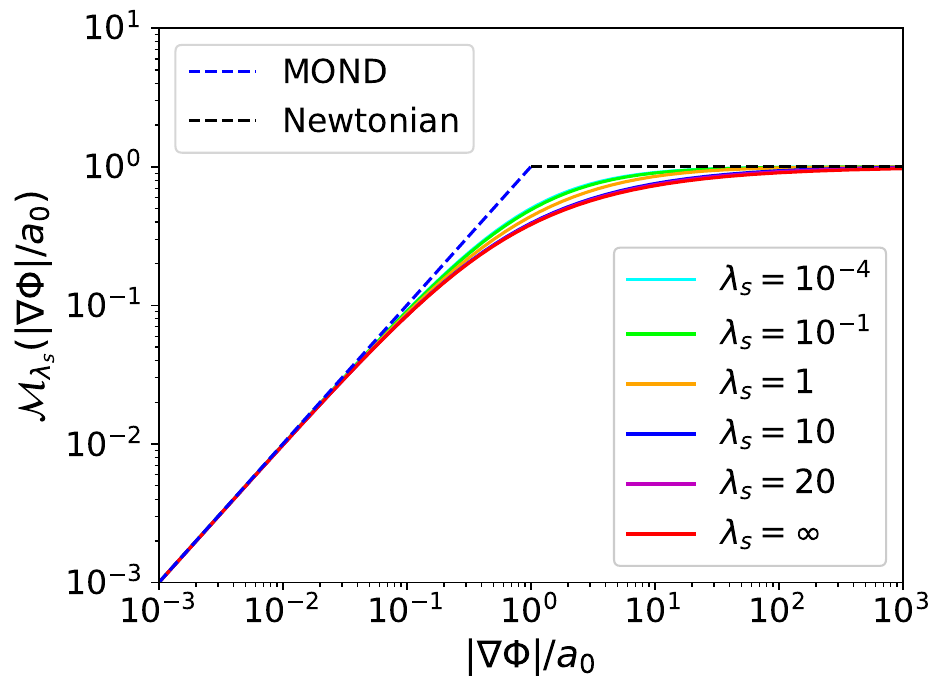}
\caption{The interpolating function $\interp_{\lambdas}(x)$ vs. $x\equiv |\grad \Phi|/a_0$ in with linear axes (left) and logarithmic axes (right) for several choices of $\lambdas$.}
\label{fig:intlinlog}
\end{figure}

In order to quantify the effects of $\lambdas$ on the interpolation function, we plot it for 
for different choices of $\lambdas$, including the case $\lambdas=\infty$ in Figure~\ref{fig:intlinlog}. We observe that as $\lambdas\rightarrow \infty$ the interpolation function has a shallower transition from MOND to Newton,
while as $\lambdas\rightarrow0$ the transition becomes somewhat sharper but not quite instantaneous. This behaviour can lead sometimes to confusion as to what exactly ``screening means''. 

\begin{figure}[H]
 \centering
 %\subfigure{}
 \includegraphics[width=0.49\textwidth]{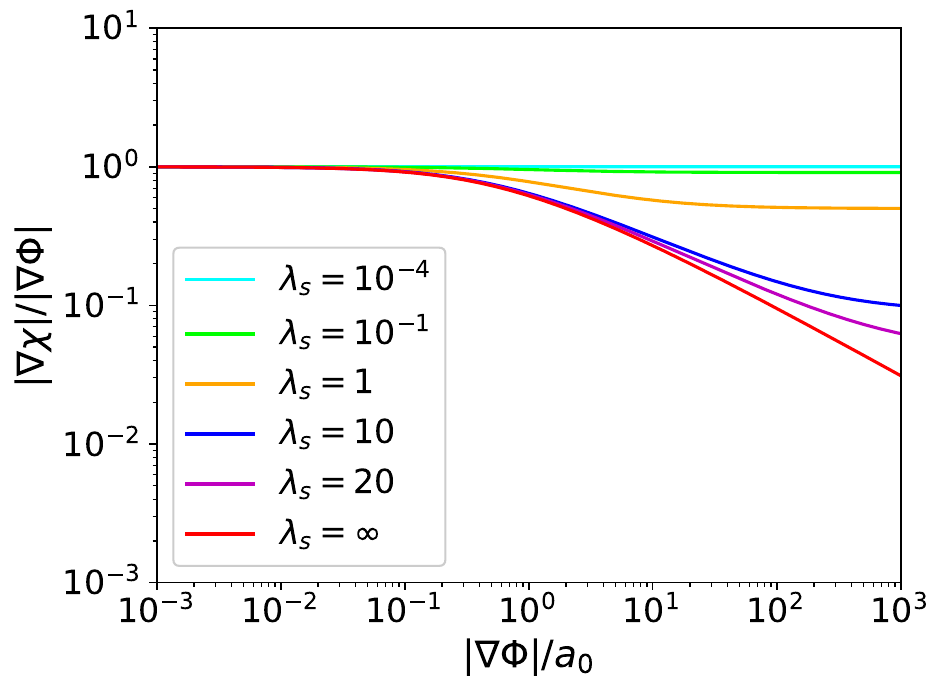}
 \caption{The ratio of the scalar field gradient to the gravitational potential gradient $|\grad \chi|/ |\grad \Phi| = {\mathcal{I}}_{\lambdas}$ vs. $x\equiv |\grad \Phi|/a_0$.}
 \label{fig:ifrac}
 \end{figure}

To quantify what is the meaning of screening, consider the ratio of the scalar field gradient $\grad \chi$ to the gravitational potential gradient $\grad\Phi$.
We have that $|\grad \chi| / |\grad \Phi| = \mathcal{I}_{\lambdas}(x) = 1 - \interp_{\lambdas}(x)$ and we plot this 
in Figure~\ref{fig:ifrac} for several choices of $\lambdas$. We observe that the $\lambdas \to \infty$ case is the case where the scalar field contributes fractionally the \emph{least} to the total gravitational potential. In other
words, as one goes to higher gravitational gradients relevant in the strong-field regime, screened cases are the ones where the scalar is mostly suppressed. The amount of suppression increases moving to higher gradients (smaller scales).
On the contrary, if $\lambdas$ is small, then this ratio tends to a constant value and therefore the scalar always contributes even as $\grad\Phi\rightarrow \infty$ --- it is not screened.

\subsection{Musings on the ``simple'' MOND interpolation function}
The simple  MOND interpolating function is defined as
\begin{align}
\interp(x) = \frac{x}{1+ x}.
\label{simple_MOND_interp}
\end{align}
Suppose that we would like to find the equivalent $\Jcal(\Ycal)$ leading to the above function, which is also screened in the deep Newtonian regime (i.e. $\beta_0 =0$). We then have that
$\reschifunc = 1 - x/(1+x) = 1/(1+x)$ so that $y(x) = x\reschifunc(x) = x/(1+x)$. This already signifies a possible issue because $y$ is bounded;
we have that $0\le y <1$, where the limit $y\rightarrow1$ is reached as $x\rightarrow \infty$. 
Still plunging ahead, $y(x)$ is a linear equation with one solution: $x(y) = y/(1-y)$, so that $\beta(y) = x(y)/y = 1/(1-y)$ leading to
$\Jcal_\Ycal(y) = y/(1-y)$. Rescaling by setting $y = \sqrt{\Ycal} / a_0$ leads to 
\begin{align}
\Jcal_{\Ycal} = \frac{\sqrt{\Ycal}}{a_0 - \sqrt{\Ycal} }
\qquad \Rightarrow 
 \qquad
\Jcal = - \Ycal -2 a_0 \sqrt{\Ycal} -2 a_0^2 \ln \left(1 - \frac{\sqrt{\Ycal}}{a_0}\right)
\label{simple_MOND_beta_0}
\end{align}  
where we have discarded the irrelevant constant of integration.
Despite appearances, $\Jcal \ge 0$ for $0\le \sqrt{\Ycal} < a_0$, and so it is expected to obey the correct stability requirements in the MOND limit found in~\cite{Skordis:2021mry}.
This function has the correct (MOND) limit as $y\rightarrow 0$ but is bounded by $\Ycal< a_0^2$ and so does not fall under our ansatz that the Newtonian 
limit is reached as $\Ycal\rightarrow \infty$. This may not be a problem, however, expanding at the limiting value by setting $\Ycal = a_0^2 (1 - \epsilon)$, 
leads to $2\ln 2 - 3  - 2 \ln \epsilon + 3\epsilon/2 + \ldots$. The $\epsilon$ term is welcome, but the logarithmic singularity $\ln \epsilon$ means that a weak-field expansion
in the limit $\Ycal \rightarrow a_0^2$ is ill-defined, and hence, the Newtonian limit (and stability requirements) are suspect and warrant further investigation.

One could try to fix the above issue with $\ln \epsilon$ by adding the term $2\Ycal  + 2 a_0^2 \ln(1 - \Ycal/a_0^2)$, which still retains a MOND limit and removes the $\ln\epsilon$
divergence. Unfortunately, this results to $\Jcal \rightarrow 2\ln 2 -1 -\epsilon/2 + \ldots$, i.e. $\epsilon$ has the wrong sign which signifies a ghost in the weak-field expansion
on a Minkowski background, in the large gradient limit (Newtonian limit). Moreover, the resulting interpolation function $\interp(x)$ is no longer  of the form $x/(1+x)$.

A final consideration is to have a non-zero $\beta_0$ in \eqref{eq:em}. After some computations we arrive to 
%\begin{align}
%  \reschifunc = 1 - \frac{1}{1+\beta_0}  \frac{x}{1+ x}
%\end{align}
%from which we get using  $ x \reschifunc(x) = \chifunct^{-1}(x) = y$ that
\begin{align}
y =  x - \frac{1}{1+\beta_0}  \frac{x^2}{1+ x} 
\end{align}
which unlike the case $\beta_0=0$, has two solutions. We choose
\begin{align}
\chifunct(y) = x(y)  =   \frac{1+\lambdas}{2} \left[-1 + y + \sqrt{   \left(1-y\right)^2+ \frac{4y}{1+\lambdas} } \right]
\end{align}
so that  from  \eqref{eq:betp} and \eqref{chifunc}   we find
\begin{align}
\Jcal_\Ycal(y) =& 
   -\frac{1-\lambdas}{2} 
+ \frac{1+\lambdas}{2y} \left[ -1 +  \sqrt{ (1  - y )^2 +  \frac{4 y}{1+\lambdas}} \right]  
% \frac{1+\lambdas}{2y} \left[-1 + y + \sqrt{ \frac{4y}{1+\lambdas} + \left(1-y\right)^2 } \right]  
\label{new_simple_MOND_beta_0}
\end{align}
where $y = \sqrt{|\Ycal|}/a_0$. This may be further integrated to get $\Jcal(\Ycal)$ analytically but we do not present its exact form here.
In the limit $\beta_0\rightarrow 0$, we recover the same functional
 form as \eqref{simple_MOND_beta_0}, provided that $y<1$. The function  \eqref{new_simple_MOND_beta_0}, however, does not have any problem in crossing the $y=1$ line if $\beta_0\ne 0$,
 and indeed $y$ can be taken to $\infty$. Moreover, we see that both $\Jcal_{\Ycal}$ and its integral $\Jcal(\Ycal)$ have 
 the correct limits (Newton and MOND) according to \eqref{def_lambdas} and \eqref{Jcal_MOND_limit}. Thus, \eqref{new_simple_MOND_beta_0} represents a one-parameter family of functions,
depending on the parameter $\beta_0$ (or equivalently $\lambdas$), from which the same interpolation function \eqref{simple_MOND_interp} emerges.

\section{Setup of numerical solutions}
\label{sec:numsol}

\subsection{Two-component system} 
\label{sec:twofc}
In this section we present the numerical procedure for the two-component system of equations and our
three cases corresponding to the density $\rho$: (i) $\rho=0$ (vacuum), (ii) $\rho(r)$ being a prescribed functional form, and (iii) $\rho$ being the solution of the hydrostatic condition \eqref{eq_rho_Phi}.
All our results were produced using our simple choice of function $\Jcal(\Ycal)$ given by \eqref{eq:simpcho} 
and we have used dimensionless variables, as defined in the relevant sections for vacuum \ref{sec:vac}, prescribed source \ref{sec:pres}, and isothermal profile \ref{sec:isoth} respectively.
We evolve the dimensionless potentials $\Phitdl$ and $\chidl$ according to \eqref{eq_Phit} and the integral of \eqref{eq_chi}, while $\Phidl = \Phitdl + \chidl$ is a derived variable.

Defining the variables  $\psi \equiv \dd\Phitdl/\drh$,    we numerically integrated the following system
\begin{align}
\frac{\dd\Phitdl}{\drh}&= \psi
\\
\frac{\dd\psi}{\drh}  &= \rhodl - \frac{2}{\rh} \psi - \muAeSTdl^2 \Phidl
\\
\frac{\dd\chidl}{\drh}  &= \frac{\psi}{\sqrt{|\psi|}}.
\end{align}
In the isothermal profile case, we define in addition the variable $\sigma \equiv -\ln\left(\rhodl\right)$ and evolve the equation
\begin{align}
\sigma' &= \left(1 + \frac{1}{\sqrt{|\psi|}}\right) \psi .
\end{align}
This system of ODEs was solved using an explicit Runge-Kutta method of order 5(4) \cite{DORMAND198019} as implemented in SciPy \cite{2020SciPy-NMeth}. We started the integration at
radius $\rh_0$ with boundary conditions $\Phitdl(\rh_{\ini}) = \Phitdl_{\ini} = 0$ and  $\chidl(\rh_{\ini}) = \chidl_{\ini}$  for the potentials, $\rhodl_{\ini}=\rhodl(\rh_{\ini})$ for the density 
(so that  $\sigma(\rh_{\ini}) = - \ln\left( \rhodl_{\ini} \right)$ in the isothermal case) and
$\psi(\rh_{\ini})=\rhodl_{\ini} \rh_{\ini}/3- \muAeSTdl^2 \Phidl_{\ini} \rh_{\ini}/3$ which arises due to the enclosed mass (gas and ) at the position $r_{\ini}$. 
 The right condition is that there is zero force at the centre. As there is a coordinate singularity we must evaluate at a position away from the centre, which necessarily encloses a tiny mass (here taken into account).

When a particular asymptotic value $\chi_\infty$ is desired, rather than having $\Phidl_{\ini}$ as boundary condition, the condition $\chidl \equiv \chidl_{\ini}$ 
can first be set to that desired value $\chidl_{\ini} = \chidl_\infty$. After one numerical iteration the integration reaches some asymptotic value $\chidl^{n}_{\ini}$ (indexed by $n$)  
and the boundary condition is updated for the next iteration so that $\Delta \chidl(\rh_{\ini}) = \chi_\infty - \chidl^n_{{\ini}}$. 
This is repeated until the asymptotic value of $\chidl$ equals $\chi_\infty$;  we found $5$ iterations usually suffice.

\subsection{Reduced Hamiltonian system}
\label{sec:redone}
In the reduced-one field case, we solve the second-order dimensionless equation for the canonical momentum $\pPhidl$ given by
\eqref{eq_P_Phi_prime_prime_vac_dl} for the vacuum case, \eqref{eq_P_Phi_prime_prime_source} for the case of prescribed source and \eqref{eq_P_Phi_prime_prime_func_iso_dl} for the isothermal case,
all corresponding to our chosen interpolation function \eqref{eq:intpfunc}. Defining $\Pi\equiv \dd\pPhidl/\drh$, The resulting equations in dimensionless form are
\begin{align}
  \frac{\dd\pPhidl}{\drh}  =&  \Pi
\\
 \frac{\dd\Pi}{\drh} =& 
\begin{cases}
  \frac{2}{\rh} \Pi   -    \muAeSTdl^2  \left[ \rh \, \sign(\pPhi) \sqrt{ |\pPhidl|  } +  \pPhidl  \right] +   \rh^2 \frac{d\rhodl_b}{ \drh} \qquad \mbox{Vacuum or prescribed source}
\\
  \frac{2}{\rh} \Pi   -   \left(   \rhodl_0  e^{\Phidl_0 - \Phidl} + \muAeSTdl^2\right)  \left[ \rh \, \sign(\pPhi) \sqrt{ |\pPhidl|  } +  \pPhidl  \right] \qquad \mbox{Isothermal}
\end{cases}
\end{align}
where in the isothermal case, $\Phidl$ is obtained from  \eqref{Wcal} below. 
The equations were solved using an explicit Runge-Kutta method of order $8(5,3)$ (DOP853) \cite{hairer2008solving} as implemented in SciPy \cite{2020SciPy-NMeth}. 
We specified boundary conditions as in the two-field system above, and translated them into boundary conditions $\hat{P}_{\Phi\ini}$ and $\Pi_{\ini}$ using
\begin{equation}
  \frac{\dd\Phidl}{\drh} =  \frac{1}{\rh} \left[  \sign(\pPhidl) \sqrt{|\pPhidl|} + \frac{\pPhidl}{\rh}  \right]
\quad \Leftrightarrow \quad
 \sqrt{|\pPhidl|}  = \frac{\rh}{2} \left( -1 + \sqrt{ 1 + 4  \left|\frac{\dd\Phidl}{\drh}\right|  } \right)
\label{relation_Phi_prime_P}
\end{equation}
which is found after combining \eqref{eq_Phi_prime} and \eqref{x_zt} for our choice of interpolation function and $\sign(\pPhidl) = \sign(\dd\Phidl / \drh)$.
Likewise the correspondence between $\Phidl$ and $\Pi$ is given by
\begin{align}
\Phidl =   \Wcal  - \frac{\Pi}{ \muAeSTdl^2  \rh^2} 
\quad \Leftrightarrow \quad
\Pi  =  \rh^2 \left( \Vcal - \muAeSTdl^2 \Phidl \right)
\label{relation_Phi_Pi}
\end{align}
where  $\Wcal  /\muAeSTdl^2 = \Vcal =\rhodl_b(\rh) $ in vacuum or for a prescribed source and 
\begin{align}
 \Wcal =  W\left[ \rhodl_0 \; e^{ \frac{ \Pi }{\muAeSTdl^2 \rh^2 } +  \Phidl_0 }  \right]  / \muAeSTdl^2 \qquad  \Vcal = \rhodl_0 e^{\Phidl_0 -  \Phidl} 
\label{Wcal}
\end{align}
for the isothermal sphere, where $W(x)$ is the Lambert (product logarithm) function.

\begin{figure}[H]
\centering
\subfigure{\includegraphics[width=0.49\textwidth]{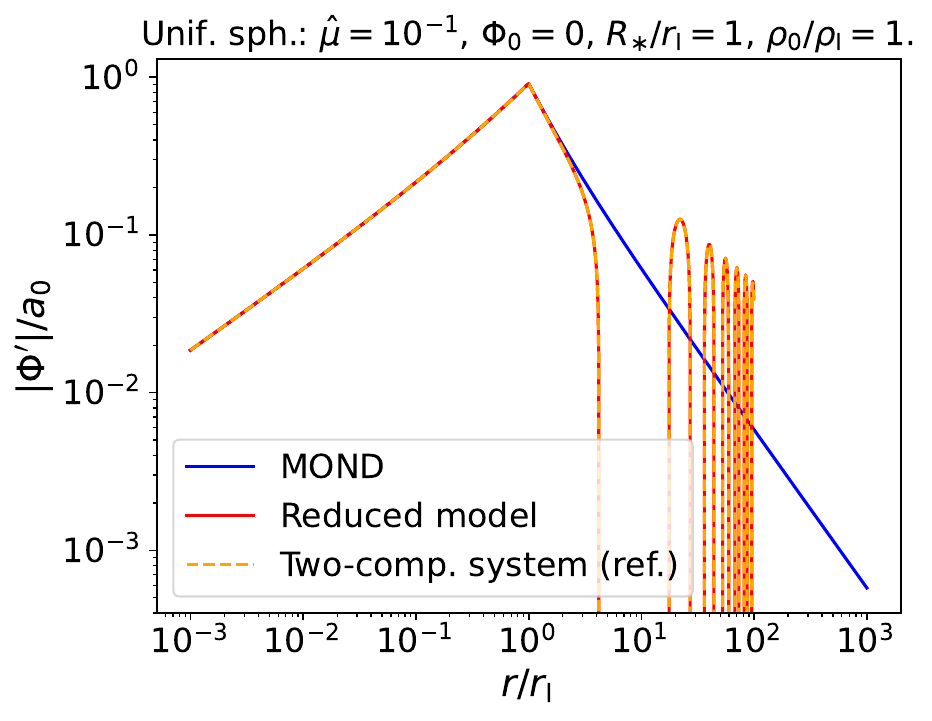}}
\subfigure{\includegraphics[width=0.49\textwidth]{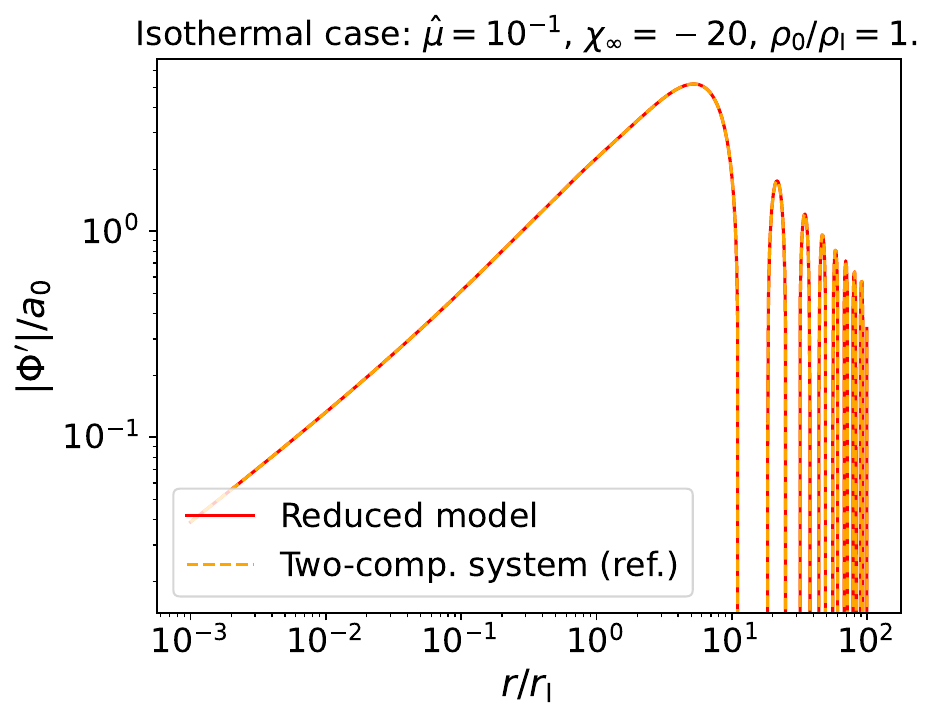}}
\caption{Comparison between the numerical solutions for $|\Phi'|/a_0$ versus dimensionless radius $\rh$ obtained with the two-component system (dashed-dotted orange line) and with the reduced Hamiltonian system (red line).
We show the case of a prescribed source (uniform density sphere, left figure) and the case of hydrostatic isothermal gas (right figure) demonstrating excellent agreement.}
\label{fig:checks}
\end{figure}

We have compared the numerical solutions of the two-component system and the reduced Hamiltonian system for the vacuum case in Figure~\ref{fig:phisols}. Here, Figure~\ref{fig:checks} shows 
the excellent agreement between the two-component system and the reduced Hamiltonian system for the uniform density sphere and for the hydrostatic isothermal gas.

\section{Low-density RAR} \label{sec:rarapp}
\begin{figure}[H]
\centering
\includegraphics[width=0.6\textwidth]{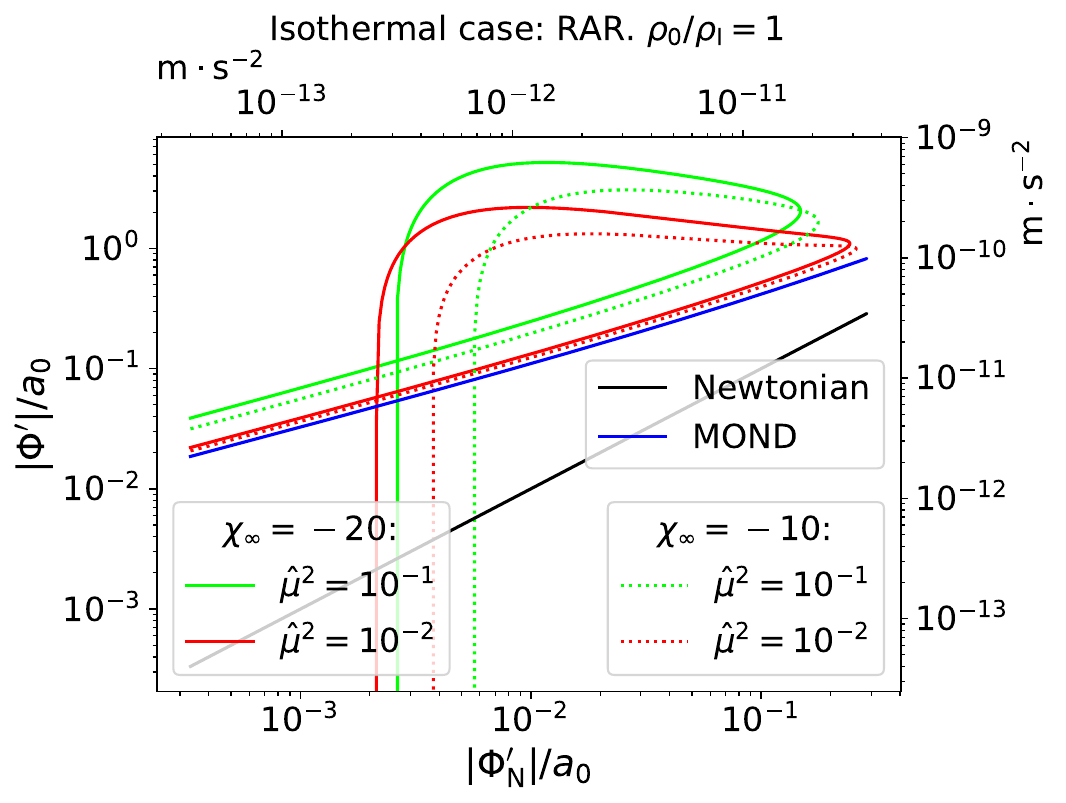}
\caption{The RAR in AeST for the hydrostatic isothermal gas when densities are low, $\rho_0/\rhoI = 1$ (in low-acceleration regime throughout). When densities are low there can be two accelerations (bifurcation) 
$\Phi'$ in AeST (seen in both red and green lines) corresponding to a given Newtonian acceleration $\Phi'_{\mathrm{N}}$, when enclosing little mass or when being farther away from the centre. This does not happen in MOND (blue line) which is one-to-one with the Newtonian acceleration.}
\label{fig:acclowd}
\end{figure}

In low-density systems it is possible for the AeST RAR to display a bifurcation, that is, have two possible values of $\Phi'$ for a given $\Phi'_{\mathrm{N}}$ as seen in Figure~\ref{fig:acclowd}.
Such a  behaviour is not possible in MOND. 

The explanation is as follows. A force may be low, and hence depart from the Newtonian expectation, for two reasons: Either little mass has been enclosed, or the distance to the centre is large. 
In MOND, the two cases are on the same point in RAR but in AeST, the presence of the mass term  $\muAeST$ distinguishes the two cases. 
It is helpful to think in terms of apparent $\chi$ condensates: One either encloses much $\chi$ condensate or little $\chi$ condensate, more $\chi$ condensate being enclosed when one is far away from the source.

\section{Pure MOND isothermal sphere}
\label{Pure_MOND_iso}

\subsection{Setting up the problem}
We would like to solve \eqref{eq_P_Phi_prime_prime_func_iso_dl_no_mu_MOND} for cases of physical interest. 
We require that the mass density is positive everywhere, implying also positivity of the mass enclosed (with $M(0) = 0$. Then \eqref{rho_pPhi_MOND} implies that $\pPhidl' \ge0$, i.e. $\pPhidl$ cannot decrease,
and moreover integrating \eqref{rho_pPhi_MOND} leads to $\Gh M(r) = \pPhi(r)$, that is $\pPhi \ge 0$ with $\pPhi=0$ only at $r=0$. 
With these considerations, \eqref{eq_Phi_prime} implies that $\Phi' \ge0$ (with strict equality only at $r=0$), and thus  \eqref{eq_rho_Phi} implies that $\rho'<0$  which means that the density will always decrease as $r\rightarrow \infty$.
Given that also $\rho\ge 0$, then $\rho$ must asymptote to a positive constant as $r\rightarrow \infty$, implying also that $\rho'\rightarrow0$ as  $r\rightarrow \infty$. Given that
$\Phi'\ne 0$ except at $r=0$, then this implies the stronger condition that $\rho\rightarrow 0$ as $r\rightarrow \infty$. From \eqref{rho_pPhi_MOND}, we then get that $\pPhidl'  \rightarrow 0$ 
as $r\rightarrow \infty$, hence, $\pPhidl$ is a monotonically increasing function which asymptotes to a constant as $r\rightarrow \infty$.

With the above considerations and since 
%Moreover, we have that shifting $\Phi$ by a constant $\Phi_0$ corresponds to rescaling of the central density $\rho_0$, leaving $\pPhidl$ invariant.
%$x = \sqrt{\zt}$
 $\rh>0$ we may set $t =  \ln \rh$ so that \eqref{eq_P_Phi_prime_prime_func_iso_dl_no_mu_MOND}  becomes an autonomous second order ODE
\begin{align}
 \frac{\dd^2 \pPhidl}{\dt^2} + (\sqrt{ \pPhidl }-3) \frac{\dd \pPhidl}{\dt} = 0
\end{align}
Now set 
\begin{align}
v =  \frac{\dd \pPhidl }{\dt}
\label{eq_v}
\end{align}
so that
\begin{align}
 \frac{\dd v}{\dt} + (\sqrt{\pPhidl}-3) v = 0
\label{eq_u}
\end{align}
and if $v\ne 0$ we combine  \eqref{eq_v} and \eqref{eq_u} to get
\begin{align}
 \frac{\dd v}{\dd \pPhidl} + \sqrt{\pPhidl }-3 = 0
\label{eq_u_vu}
\end{align}
which solves to
\begin{align}
 v = 9 v_0 + 3\pPhidl  - \frac{2}{3} \pPhidl ^{3/2}
\end{align}
where $v_0$ is a constant.
Thus, from \eqref{eq_v} we find the first order separable ODE
\begin{align}
  \frac{\dd \pPhidl }{\dt} = 9 v_0 + 3\pPhidl  - \frac{2}{3}\pPhidl^{3/2}
\label{eq_v_0}
\end{align}
which integrates to
\begin{align}
 t = t_0 + \int  \frac{\dd \pPhidl }{ 9 v_0 + 3\pPhidl  - \frac{2}{3}\pPhidl^{3/2}   } .
\end{align}
We require that $\pPhi\ge0$  and $\pPhi'\ge 0$. If $v_0=0$, then $0\le \pPhidl  \le \frac{81}{4}$ with $\pPhidl(0)=0$ and $\pPhidl^{(\infty)}=\pPhidl(\rh\rightarrow \infty) = 81/4$. 
Increasing $v_0$ has no impact on the lower boundary while $\pPhidl^{(\infty)}$ increases to higher values. However, negative $v_0$, introduces a minimum $\pPhidl$ which cannot happen at $\rh=0$. Thus for 
the $v_0<0$  to be physical, they must connect to a new regime (e.g. Newtonian) which should be valid at small radii down to $\rh=0$. For $v_0\ge-1$,  $\pPhi'<0$ except at a point, hence,
 there are no physical solutions of any kind in this case.

Now let us perform the integral. Since $\pPhidl>0$ by definition, we set $\pPhidl = (q + 3/2)^2$, with $q\ge -3/2$, so that
\begin{align}
 t = t_0 - 3\int  \frac{q + \frac{3}{2}}{ q^3 - \frac{27}{4}q  + \tilde{v}_0 }  \dd q
\end{align}
where $\tilde{v}_0 =  - \frac{27}{4}\left(1 + 2v_0 \right) $. We need to factorize the denominator so that we can expand the integrand in partial fractions
and for this we need the roots of the cubic $ q^3 - \frac{27}{4}q  + \tilde{v}_0  =0$. Clearly $\tilde{v}_0$, and thus $v_0$, play a fundamental role in how many real roots there can be. 
The discriminant  of the cubic is $\Delta_q = - \frac{729}{4} v_0 (1+v_0)$ and if $v_0>0$ then there is only one real root, $q_1$, given by~\cite{Holmes2002} 
\begin{align}
 q_1  = 3  \cosh\left[ \frac{1}{3} \arccosh\left( 1 + 2v_0 \right) \right]
\end{align}
and we have that $q_1>3$.  If $v=0$ then the roots are $q_I=\{3,-\frac{3}{2},-\frac{3}{2}\}$ and for $-1<v_0<0$ there are three real roots $q_I$.

\subsection{The case $v_0=0$}
The case $v_0=0$ can be easily integrated as we have
\begin{align}
 t = t_0 - 3\int  \frac{1}{(q-3)(q+\frac{3}{2})}  \dd q = t_0  - \frac{2}{3} \ln\frac{\abs{q-3}}{\abs{q+\frac{3}{2}}} 
\end{align}
leading after changing back to the original variables
\begin{align}
\pPhidl =& \frac{81}{4} \frac{\rh^3/\rh_0^3}{\left(\sqrt{ \rh^3/\rh_0^3} + 1\right)^2 } \qquad \forall \quad \rh>0
\label{Core_ISO_1}
\end{align}
and
\begin{align}
\pPhidl =& \frac{81}{4} \frac{ \rh^3/\rh_0^3 }{\left(\sqrt{ \rh^3/\rh_0^3} - 1\right)^2 } \qquad \forall \quad \rh> \rh_0.
\label{Sing_ISO_1}
\end{align}
Solution \eqref{Core_ISO_1} is a cored isothermal sphere valid for all $\rh>0$ while solution
\eqref{Sing_ISO_1} is a singular isothermal sphere valid for $\rh>  \rh_0$, with the singularity occurring exactly at the lower boundary.

\subsection{The case $v_0>0$}
In the case $v_0>0$, $\pPhi$ takes values in the range $0\ge \pPhi \ge \pPhidl^{(\infty)}$ where $\pPhidl^{(\infty)}>81/4$. At $\rh=0$, the derivative $\pPhi' > 0$ is strictly positive.
 We write the integral as
\begin{align}
 t =& t_0 - 3\int  \frac{q + \frac{3}{2}}{ (q-q_1) \left(q^2 + q_1 q   + q_1^2 - \frac{27}{4}\right) }  \dd q
\\
=& t_0
+  \ln \left( \frac{ \sqrt{ q^2+ q_1q + q_1^2 - \frac{27}{4}}}{ \abs{q-q_1}  }\right)^{\frac{1}{q_1-\frac{3}{2}}  }
- \frac{1}{q_1 - \frac{3}{2}} \sqrt{3 \frac{ q_1-3}{ q_1+3 }} \arctan\left(\frac{2 q+q_1}{\sqrt{3} \sqrt{q_1^2-9}}\right)
\label{eq_case_v_0}
\end{align}
where we have used that $q^2 + q_1 q   + q_1^2 - \frac{27}{4}\ge 0$ for $q\ge-3/2$, and 
where the second line follows after expanding the integrand in partial fractions, and integrating some of them to get the logarithms while completing a square in one term, leading to $\arctan$. Back to the original variables 
we find a branch $0\le \pPhi \le \pPhi^{(\infty)}$ (corresponding to $-3/2\le q \le q_1$) with $\pPhi^{(\infty)}  = (q_1 + 3/2)^2$, which is the analogue of \eqref{Core_ISO_1} and valid for $\rh\ge 0$,
and a branch $\pPhi \ge \pPhi^{(\infty)}$ which is the analogue of \eqref{Sing_ISO_1} and valid for $\rh\ge \rh_0$. Unfortunately, \eqref{eq_case_v_0} cannot be inverted and find
$\pPhi(r)$ in terms of elementary functions, however, we can consider the large $r$ expansion of $\pPhi$.
%, that is,
%case $\pPhi  = \epsilon$ and the case 
%$\pPhi = \pPhi^{(\infty)} - \epsilon $, as $\epsilon \rightarrow 0$.
%The first case (small $r$)  leads to
%\begin{align}
%\pPhidl \approx&
%\end{align}
%while the second case (large $r$) leads to
We find
\begin{align}
\pPhidl \approx& \left(q_1 + \frac{3}{2} \right)\left\{ q_1 + \frac{3}{2}    - 2 \left[ \left(\frac{81}{4}\right)^{1/3} \frac{\rh_0}{\rh}\right]^{q_1-\frac{3}{2}} + \ldots
\right\}.
\label{pPhi_large}
%\underset{q_1\rightarrow3}{\rightarrow}\frac{81}{4}  - \frac{81}{2} \left(\frac{r_0}{\rh} \right)^{3/2}  + \ldots
\end{align}
Considering  \eqref{pPhi_large} further we find the large $r$ limit of the density
\begin{equation}
4\pi G \rho \approx \frac{4}{9} \left(\frac{9}{2}\right)^{\frac{2q_1}{3}} \frac{a_0^2 \xi\left(q_1^2 - \frac{9}{4}\right)}{\rh_0^3} \left(\frac{\rh_0}{a_0 \xi r}\right)^{q_1 + \frac{3}{2} }
\end{equation}
potential
\begin{equation}
\Phidl  \approx  \Phidl_0
+ \frac{1}{\xi} \ln\left[ \frac{9\rhodl_0 \rh_0^3}{4q_1^2 - 9} \left(\frac{2}{9}\right)^{\frac{2q_1}{3}} \right]
 + \frac{q_1 + \frac{3}{2}}{\xi} \ln\left(\frac{\rh}{\rh_0}\right)
\end{equation}
and force
\begin{equation}
\frac{\dd\Phidl}{\drh} \approx \frac{q_1 + \frac{3}{2}}{\xi} \frac{1}{\rh}.
\end{equation}
In the limit $q_1\rightarrow 3$ (i.e. $v_0\rightarrow 0$), all approximations above agree with the expansion of \eqref{Core_ISO_1} for small and large $r$ respectively.
%\begin{align}
% \frac{\dd^2\epsilon}{\drh^2}  =&  
%+\frac{2}{\rh} \frac{\dd\epsilon}{\drh}  
%-  \rh  \frac{4}{9} \left(\frac{9}{2}\right)^{\frac{2q_1}{3}} \frac{a_0^2 \xi\left(q_1^2 - \frac{9}{4}\right)}{\rh_0^3} \left(\frac{\rh_0}{a_0 \xi r}\right)^{q_1 + \frac{3}{2} }
%  \sqrt{  \pPhidl_M } \frac{\epsilon}{ 2 \pPhidl^M  }  
%-  \rh \muAeSTdl^2 \sqrt{  \pPhidl^M } 
%\end{align}

\end{document}